\documentclass[pre,twocolumn,showpacs,longbibliography]{revtex4-1}

\newcommand{\cP}{{\cal P}}
\newcommand{\cPmax}{\cP_p}

\newcommand{\Eq}[1]{Eq.~(\ref{eq:#1})}
\newcommand{\Eqs}[2]{Eqs.~(\ref{eq:#1}) and~(\ref{eq:#2})}
\newcommand{\f}{\mathbf{f}}
\newcommand{\Fig}[1]{Fig.~\ref{fig:#1}}
\newcommand{\Figure}[1]{Figure~\ref{fig:#1}}
\newcommand{\Figures}[1]{Figures~\ref{fig:#1}}

\newcommand{\REF}[1]{Ref.~\cite{#1}}

\newcommand{\Sec}[1]{Sec.~\ref{sec:#1}}
\newcommand{\App}[1]{Appendix~\ref{sec:#1}}
\renewcommand{\f}{\mathbf{f}}
\newcommand{\fel}{f^\mathrm{el}}
\newcommand{\rr}{\mathbf{r}}
\renewcommand{\v}{\mathbf{v}}

\newcommand{\expt}[1]{\left< #1\right>}
\newcommand{\gdot}{\dot{\gamma}}

\newcommand{\Tab}[1]{Table~\ref{tab:#1}}

\newcommand{\vtot}{v^\mathrm{tot}}
\usepackage{graphicx}
\usepackage{amssymb}
\usepackage{amsmath}
\usepackage{bm}

\begin{document}
\title{Slow and fast particles in shear-driven jamming: critical behavior}

\author{Peter Olsson}

\affiliation{Department of Physics, Ume\aa\ University, 
  901 87 Ume\aa, Sweden}

\date{\today}   

\begin{abstract}
  We do extensive simulations of a simple model of shear-driven jamming in two dimensions
  to determine and analyze the velocity distribution at different densities $\phi$ around
  the jamming density $\phi_J$ and at different low shear strain rates, $\gdot$. We then
  find that the velocity distribution is made up of two parts which are related to two
  different physical processes which we call the slow process and the fast process as they
  are dominated by the slower and the faster particles, respectively. Earlier scaling
  analyses have shown that the shear viscosity $\eta$, which diverges as the jamming
  density is approached from below, consists of two different terms, and we present strong
  evidence that these terms are related to the two different processes: the leading
  divergence is due to the fast process whereas the correction-to-scaling term is due to
  the slow process. The analysis of the slow process is possible thanks to the observation
  that the velocity distribution for different $\gdot$ and $\phi$ at and around the
  shear-driven jamming transition, has a peak at low velocities and that the distribution
  has a constant shape up to and slightly above this peak. We then find that it is
  possible to express the contribution to the shear viscosity due to the slow process in
  terms of height and position of the peak in the velocity distribution and find that this
  contribution matches the correction-to-scaling term, determined through a standard
  critical scaling analysis.  A further observation is that the collective particle motion
  is dominated by the slow process. In contrast to the usual picture in critical phenomena
  with a direct link between the diverging correlation length and a diverging order
  parameter, we find that correlations and shear viscosity decouple since they are
  controlled by different sets of particles and that shear-driven jamming is thus an
  unusual kind of critical phenomenon.
\end{abstract}

\maketitle

\section{Introduction}

Particle transport is an ubiquitous phenomenon with relevance for both industry and
every-day life and the behaviors of such real-life systems are immensely complicated as
they include effects of e.g.\ varying particle shape, friction, and gravity. Even
idealized systems \cite{OHern_Silbert_Liu_Nagel:2003} where such complications can be
eliminated---spherical (or circular) particles without any friction and well-controlled
volume or pressure---remain poorly understood. Some salient features are that the shear
viscosity increases as the packing fraction $\phi$ approaches the jamming packing fraction
$\phi_J$ from below, that the relaxation time increases, and that the particle motion
becomes increasingly correlated. It has however been difficult to find a way to connect
together different quantities and behaviors into a comprehensive picture.

Simulations of shear-driven jamming are typically performed at constant packing fraction
$\phi$ and low shear strain rates $\gdot$ \cite{Durian:1995}, and some of the quantities
of interest are pressure $p$ and shear stress $\sigma$.  One important characterization of
the shear-driven jamming transition is through the value of the critical exponent $\beta$
that describes the divergence of the shear viscosity, $\eta\equiv\sigma/\gdot$, as the
jamming density $\phi_J$ is approached from below,
\begin{equation}
  \label{eq:eta-dphi}
  \eta \sim (\phi_J-\phi)^{-\beta}.
\end{equation}
A starting point for many theoretical attempts to understand shear-driven jamming has been
properties of static jammed packings at, or slightly above, jamming. A collection of
particles with contact-only interactions forms a rigid network just at the jamming
transition, with the number of contacts per particle equal to $z=z_c\equiv 2d$
\cite{Alexander:1998}, (with the generalization to a finite number of particles in
\REF{Goodrich:2012}), and both the distance between close particles and the weak contact
forces for contacting particles follow power-law distributions with non-trivial exponents
\cite{Charbonneau:NatCommun:2014, Charbonneau:Jstat:2014, Charbonneau:2015:prl}. From the
values of these exponents, expected to be the same for dimension $d\geq2$, together with
some additional assumptions, one has found $\beta/u_z\approx 3.41$ \cite{DeGiuli:2015,
  Harukuni-logcorr:2020} for the exponent that describes the dependence of the viscosity
on the distance to isostaticity, $\eta\sim(z-z_c)^{-\beta/u_z}$. This may be compared with
results from simulations in two dimensions that have generally given lower values:
$\beta/u_z=1/0.38=2.63$ \cite{Lerner-PNAS:2012} and $\beta/u_z=2.69$
\cite{Olsson:jam-tau}. (A later work by the group of \REF{Lerner-PNAS:2012} gave a higher
value, $\beta/u_z\approx 3.3$ \cite{DeGiuli:2015}, in agreement with the theoretical
value, but that was for three dimensions ; determinations in two dimensions tend to give
lower values \cite{Olsson:jam-3D, Nishikawa_Ikeda_Berthier:2021, Olsson:jam-NIB}.)
Similarly, the values of $\beta$ in two dimensions, which have typically been in the range
$\beta=2.2$ through 2.83 \cite{Andreotti:2012, Olsson_Teitel:gdot-scale,
  Kawasaki_Berthier:2015} are found to be in agreement with the lower values
($\beta/u_z\approx 2.69$) when using $u_z=1$ \cite{Heussinger_Barrat:2009}. One way to
explain this discrepancy between the theoretically found $\beta/u_z\approx 3.41$
\cite{DeGiuli:2015, Harukuni-logcorr:2020} and the lower values from simulations is to
claim that these lower values are incorrect due to a neglect of logarithmic corrections to
scaling \cite{Harukuni-logcorr:2020}. This is a possibility since the upper critical
dimension of the jamming transition is widely believed to be $d_\mathrm{ucp}=2$
\cite{Wyart:2005, Goodrich:2012}, which opens up for logarithmic corrections to
scaling. Though this explanation is a possibility, it could also be that the discrepancy
only points to a lack of understanding of the phenomenon of shear-driven jamming.

Of the mentioned works, \REF{Lerner-PNAS:2012} from simulations of hard disks, and the
simulations that are based on relaxing configurations of soft disks below $\phi_J$
\cite{Olsson:jam-tau, Olsson:jam-3D, Nishikawa_Ikeda_Berthier:2021, Olsson:jam-NIB},
determine the divergence in terms of $\delta z$, and do not give any value for
$\phi_J$. The other works mentioned above are from simulations of soft disks
\cite{Andreotti:2012, Olsson_Teitel:gdot-scale, Kawasaki_Berthier:2015} and rely on
scaling relations in one way or the other.

It has long been realized that the particle motion becomes increasingly collective as
$\phi_J$ is approached from below \cite{Pouliquen:2004}. One way to study this in
simulations is with the overlap function \cite{Lechenault_2008,
  Heussinger_Berthier_Barrat:2010} and the associated dynamic susceptibility, $\chi_4$,
which gives a measure of the number of particles that move collectively. With the
assumption that the correlated domains have a compact geometry that quantity gave a length
diverging with $\nu=0.9$; similar exponents were found also from other quantities
\cite{Heussinger_Berthier_Barrat:2010}. From a correlation function that, in contrast to
$\chi_4$, makes use of the vectorial nature of the velocity field, it has also been found
that it is possible to extract two correlation lengths from the velocity fluctuations,
respectively related to the rotation and the divergence of the velocity field. It appears
that it is the length scale related to the rotations that is the more significant one
\cite{Olsson_Teitel:jam-xi-ell}.

With a diverging length scale and a diverging dynamic quantity, $\eta$, it could seem that
the jamming transition fits nicely into the ordinary description of a critical
phenomenon. It has however been difficult to understand the detailed connection between
these two quantities. The divergence of the correlation length with $\nu=1$ has sometimes
been taken to suggest $\beta=2$---one way to get that result is from the derivation of
\Eq{eta_slow} in \Sec{ratio} below---which is difficult to reconcile with the range of
$\beta$ values given above.

In this paper we present evidence for, and explore some consequences of, the existence of
two different processes in the system with different scaling properties: the \emph{fast
  process} which is dominated by fast particles from the tail of the velocity distribution
and the \emph{slow process} which is dominated by the big fraction of slow particles from
the peak of the distribution. It has already been shown that the divergence of the
viscosity is dominated by a small fraction of particles with the highest velocities
\cite{Olsson:jam-vhist}, which means that the behavior described in \Eq{eta-dphi} is
controlled by the fast process.  In this paper we show that the collective motion is
governed by the slow process. A consequence is that the link between correlation length
and the diverging shear viscosity is only an indirect one, which seems to imply that
shear-driven jamming is a very unusual kind of critical phenomenon.

The analyses in the presented paper are for two-dimensional systems, only. Preliminary
studies in three and four dimensions do however show that the same kind of analysis works
very well also in these higher dimensions, and we therefore expect the conclusions to hold
also in the more physically relevant case of three dimensions. These results will be
presented elsewhere.

Though a critical divergence of a quantity as in \Eq{eta-dphi} is described by a critical
exponent there are usually additional terms that have to be included in the analyses
unless one happens to have access to data only very close to the critical point. This goes
under the heading of ``corrections to scaling'' and is due to the presence of irrelevant
variables in the scaling function. In shear-driven jamming one has indeed found that a
single diverging term cannot successfully fit the data \cite{Olsson_Teitel:gdot-scale,
  Kawasaki_Berthier:2015} and the inclusion of a correction-to-scaling term was found to
give reasonable analyses. The finding of two different processes in shear-driven jamming,
however, opens up for a different interpretation of this additional term. The evidence
suggests that the correction-to-scaling term is due to the slow process which means that
it is possible to relate this term to a separate physical process, which is unusual for
critical phenomena.

The remainder of the paper is organized as follows: In Sec.~II we describe the simulations
and the measured quantities and give a motivation for the use of the velocity distribution
for analyzing shear-driven jamming. We also review the scaling relations and discuss
shortly different ways to analyze the transition.  In Sec.~III we describe the results, to
a large extent through analyses of data at $\phi\approx\phi_J$. We do this by first
showing that the correction-to-scaling term of the shear stress may be related to the
properties of the peak in the velocity distribution. We then first turn to the behavior at
densities in a (narrow) interval around $\phi_J$ and show that the two different
terms---where one is the contribution to $\sigma$ from the peak in the distribution and
the other is the remainder---both scale with $\phi-\phi_J$ and $\gdot$. We then also show
that the same kind of analysis may actually be used also in the hard disk limit, i.e.\ in
the region well below $\phi_J$ and at sufficiently low $\gdot$ that the shear viscosity is
independent of shear rate.  We also discuss the origin of the high velocities of the fast
process and then turn to the collective particle motion and argue that the diverging
correlation length and the leading divergence of the shear viscosity, as jamming is
approached, are due to different sets of particles. We then present a rationalization of
some of our findings. In Sec.~IV we finally summarize the results, discuss some open
questions and some connections between our findings and the literature, and sketch a few
directions for future research.

A jointly published Letter \cite{jointPRL} summarizes some of our key results. The Letter
also shortly discusses finite size scaling, which will be discussed in more detail in a
separate publication.

\section{Models and measured quantities}
\label{sec:model}

\subsection{Simulations}

For the simulations we follow O'Hern \emph{et al.} \cite{OHern_Silbert_Liu_Nagel:2003} and
use a simple model of bi-disperse frictionless disks in two dimensions with equal numbers
of particles with two different radii in the ratio 1.4. We use Lees-Edwards boundary
conditions \cite{Evans_Morriss} to introduce a time-dependent shear strain
$\gamma = t\gdot$. With $r_{ij}$ the distance between the centers of two particles and
$d_{ij}$ the sum of their radii, the relative overlap is $\delta_{ij} = 1 - r_{ij}/d_{ij}$
and the interaction between overlapping particles is
$V_p(r_{ij}) = \epsilon \delta_{ij}^2/2$; we take $\epsilon=1$. The force on particle $i$
from particle $j$ is $\f^\mathrm{el}_{ij} = -\nabla_i V_p(r_{ij})$, which gives the force
magnitude $f^\mathrm{el}_{ij}=\epsilon\delta_{ij}/d_{ij}$. The total elastic force on a
particle is $\f^\mathrm{el}_i = \sum_j \f^\mathrm{el}_{ij}$ where the sum is over all
particles $j$ in contact with $i$.

The simulations discussed here have been done at zero temperature with the RD$_0$
(reservoir dissipation) model \cite{Vagberg_Olsson_Teitel:BagnNewt} with the dissipating
force $\f^\mathrm{dis}_i = -k_d \v_i$ where $\v_i\equiv \v_i^\mathrm{tot}-y_i\gdot\hat x$
is the non-affine velocity, i.e.\ the velocity with respect to a uniformly shearing
velocity field, $y_i\gdot\hat x$.  In the overdamped limit the equation of motion is
$\f^\mathrm{el}_i +\f^\mathrm{dis}_i = 0$ which becomes $\v_i = \f^\mathrm{el}_i/k_d$.  We
take $k_d=1$ and the time unit $\tau_0 = d_s^2 k_d/\epsilon=1$. Length is measured in
units of the diameter of the small particles, $d_s$. The equations of motion were
integrated with the Heuns method with time step $\Delta t/\tau_0=0.2$.  Unless otherwise
noted the results are for $N=65536$ particles.

\subsection{Measured quantities}

Using $\rr_{ij} = \rr_i - \rr_j$ we determine the pressure tensor,
\begin{equation}
  \mathbf{p}^\mathrm{el} = \frac{1}{V}\sum_{i<j} \f^\mathrm{el}_{ij}\otimes\rr_{ij},
  \label{eq:ptensor}
\end{equation}
which is measured during the simulations once per unit time.  Here $V=L\times L$ is the
volume. The pressure is obtained from the pressure tensor through
\begin{displaymath}
  p=\frac 1 2 [\expt{\mathbf{p}^\mathrm{el}_{xx}} + \expt{\mathbf{p}^\mathrm{el}_{yy}}],
\end{displaymath}
and the shear stress is given by
\begin{equation}
  \label{eq:sigma}
  \sigma = - \expt{\mathbf{p}^\mathrm{el}_{xy}}.
\end{equation}

The analyses below will focus on the dissipation and a crucial relation is then the
connection between shear stress and $\expt{v^2}$ (where $v\equiv|\v|$ is the non-affine
velocity) which follows from the requirement of power balance between the input power
$V\sigma\gdot$ and the dissipated power $k_d\sum v_i^2$, where the sum is over all the
particles.  This gives \cite{Ono_Tewari_Langer_Liu}
\begin{equation}
  \label{eq:power}
  \sigma\gdot = \frac N V k_d \expt{v^2},
\end{equation}
which implies $\sigma \sim \expt{v^2}/\gdot$.

\subsection{The velocity distribution}

Though \Eq{power} could lead to the thinking that measures of the velocity and measures of
$\sigma$ only give the same information, our claim is there is more information in the
velocity distribution.  To see this we consider the behavior of continuously sheared hard
spheres below $\phi_J$. For that case it has been found that the displacement (i.e.\
velocity) is governed by steric exclusion \cite{Andreotti:2012} and that the forces at
each moment will adjust to give the velocities that are required by steric hindrance. This
implies that the forces and the shear stress are controlled by the velocity and it also
suggests that velocity is a more fundamental quantity, and that there might be more
information in the full velocity distribution than what is contained in the shear stress,
$\sigma$. In the present work we set out to extract some of that information.

To measure the distribution function $\cP(v)$ we define the bin size $\Delta$ and
$v_k=k\Delta$ and let the histogram $H(v_k)$ be the fraction of the non-affine particle
velocities in the range $[v_k-\Delta/2,v_k+\Delta/2)$. Histograms are created from files
with configurations that are stored every 10~000 time step.  The distribution function,
$\cP(v_k)=H(v_k)/\Delta$, is normalized such that $\int \cP(v) dv=1$.  From \Eq{power}
follows an expression for the shear stress in terms of the velocity distribution function,
\begin{equation}
  \label{eq:sigma-intv}
  \sigma = \frac{N}{V} \frac{k_d}{\gdot}\int \cP(v) v^2 dv.
\end{equation}

\subsection{Scaling relations}

For easy reference we here show derivations of some scaling relations from the standard
scaling assumption \cite{Olsson_Teitel:jamming,  Vagberg_Olsson_Teitel:CDn},
\begin{equation}
  \label{eq:sigma-b}
  \sigma(\phi, \gdot)b^{y/\nu} =  \bar g_\sigma(\delta\phi\, b^{1/\nu}, \gdot b^z)
  + b^{-\omega} \bar h_\sigma(\delta\phi\, b^{1/\nu}, \gdot b^z).
\end{equation}
Here $b$ is a length rescaling factor, $y$ is the scaling dimension of $\sigma$, $\nu$ is
the correlation length exponent, $\delta\phi=\phi-\phi_J$, $z$ is the dynamical exponent,
$\omega$ is the correction-to-scaling exponent and $\bar g_\sigma$ and $\bar h_\sigma$
are unknown scaling functions.

With $b=\gdot^{-1/z}$ in \Eq{sigma-b} and with $q=y/z\nu$ one finds
\begin{equation}
  \label{eq:sigma-scale}
  \sigma(\phi,\gdot) = \gdot^q\left[g_\sigma\left(\frac{\phi-\phi_J}{\gdot^{1/z\nu}}\right)
    + \gdot^{\omega/z} h_\sigma\left(\frac{\phi-\phi_J}{\gdot^{1/z\nu}}\right) \right].
\end{equation}
One way to determine the critical behavior of the shear-driven jamming transition has been
to fit $\sigma(\phi,\gdot)$ or $p(\phi,\gdot)$ at densities around $\phi_J$, to this
expression \cite{Olsson_Teitel:gdot-scale}. The scaling functions $g_\sigma$ and
$h_\sigma$ were there taken to be exponentials of polynomials in
$(\phi-\phi_J)/\gdot^{1/z\nu}$, and both $\phi_J$ and the critical exponents were
determined through scaling fits of both $p$ and $\sigma$.

Right at $\phi_J$, with the notation $q_2=q+\omega/z$, \Eq{sigma-scale} becomes
\begin{equation}
  \label{eq:sigma-two}
  \sigma(\phi_J,\gdot) = \gdot^q g_\sigma(0) + \gdot^{q_2} h_\sigma(0).
\end{equation}
The conclusion of a behavior as in \Eq{sigma-two} was reached
in a different way in \REF{Kawasaki_Berthier:2015}.  An analysis, consistent with
\Eq{sigma-two}, of a similar model, commonly used for granular materials, has also been
done \cite{Rahbari_Vollmer_Park}.

To get the scaling relation for the shear viscosity one writes an expression for
$\sigma(\phi, \gdot)b^{y/\nu}/(\gdot b^z)$ from \Eq{sigma-b} and takes
$b=(-\delta\phi)^{-\nu}$. This then becomes
\begin{eqnarray}
  \label{eq:eta-gdot}
  \eta(\phi, \gdot) & = & (\phi_J-\phi)^{-\beta}
                          g_\eta\left(\frac{\gdot}{(\phi_J-\phi)^{z\nu}}\right) + \nonumber \\
  & + & (\phi_J-\phi)^{-\beta_2} h_\eta\left(\frac{\gdot}{(\phi_J-\phi)^{z\nu}}\right),
\end{eqnarray}
where $\beta=z\nu-y$ and $\beta_2=z\nu-y-\omega\nu$. The first term is the leading
divergence and the second is the correction to scaling term. When comparing with the
expressions for $q$ and $q_2$ in \Eq{sigma-two} one finds
\begin{subequations}
  \label{eq:beta_beta2}
  \begin{eqnarray}
    \label{eq:beta}
    \beta/z\nu & = & 1-q,\\
    \label{eq:beta2}
    \beta_2/z\nu & = & 1-q_2.
  \end{eqnarray}
\end{subequations}
For sufficiently small $\gdot$ the scaling functions in \Eq{eta-gdot} approach constants,
and one arrives at
\begin{equation}
  \label{eq:eta-with-corr}
  \eta(\phi,\gdot\to0) = c_1 (\phi_J-\phi)^{-\beta} + c_2 (\phi_J-\phi)^{-\beta_2},
\end{equation}
which is the behavior in the hard disk limit.

One approach to shear-driven jamming is then to consider the shearing of a collection of
hard disks (or soft disk in the limit $\gdot\to0$) below $\phi_J$, and thus with the
average number of contacts $z<z_c$. This is sometimes called the floppy flow
regime. Another approach, relevant at higher shear strain rates and/or closer to $\phi_J$,
is to examine the behavior where the elasticity of the particles is important. This is the
elasto-plastic regime which at $\phi_J$ is described by \Eq{sigma-two}. Though it could
seem that the behaviors in these different regimes are governed by very different physical
processes, we note that the respective behaviors both follow from a single scaling
assumption, which suggests that both regions are governed by the same fundamental physics.

The present article presents a novel analysis of the shear-driven jamming transition. Most
of the analyses are done on data at $\phi=\phi_J$ and for different $\gdot$, but in
\Sec{hard-particle-limit} we demonstrate that the same kind of analysis works well also
for data in the hard disk limit at $\phi<\phi_J$.

\section{Results}

\subsection{Two terms in $\sigma$}
\label{sec:Two-sigma}

\begin{figure}
  \includegraphics[width=8cm]{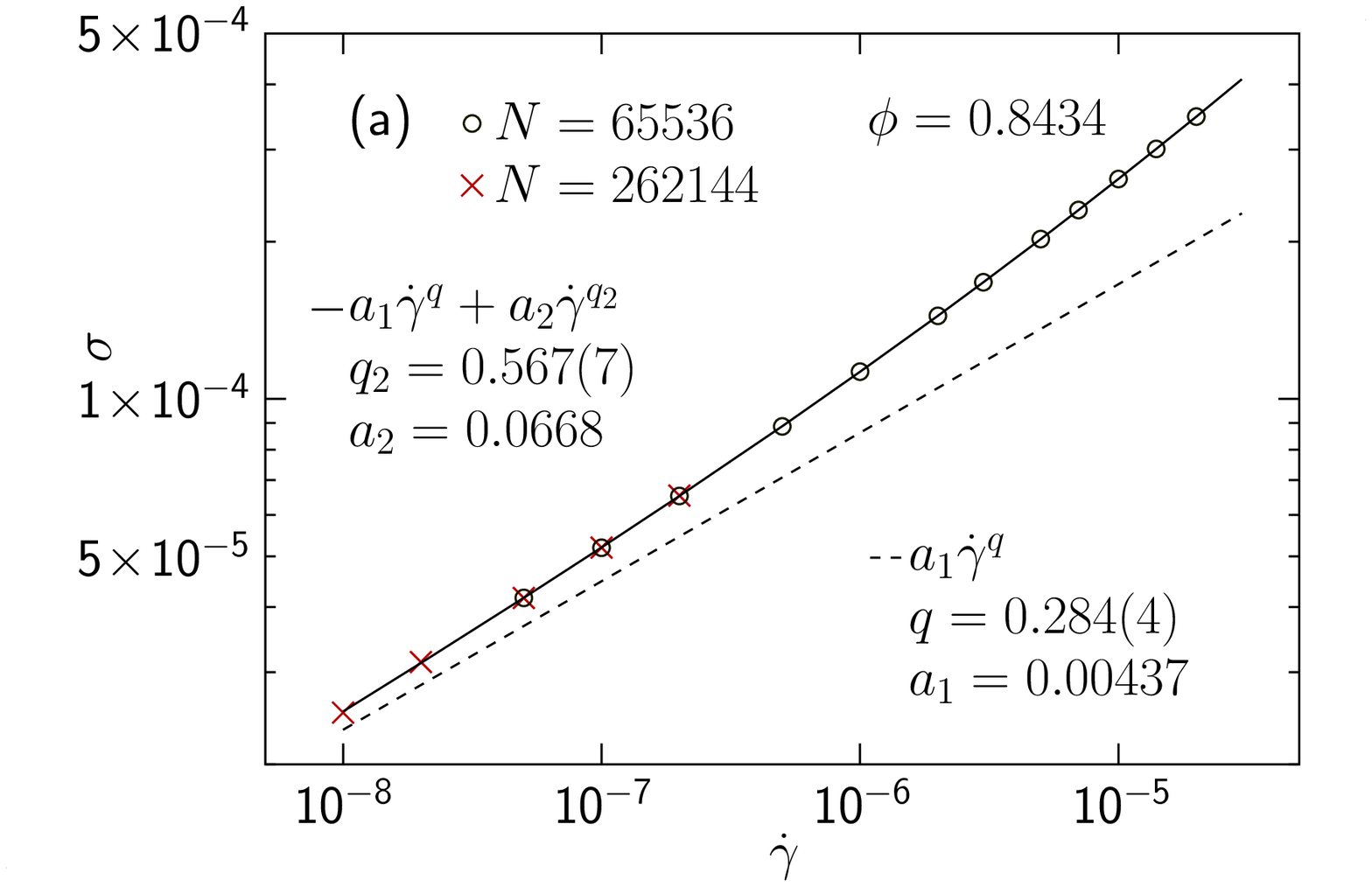}
  \includegraphics[width=8cm]{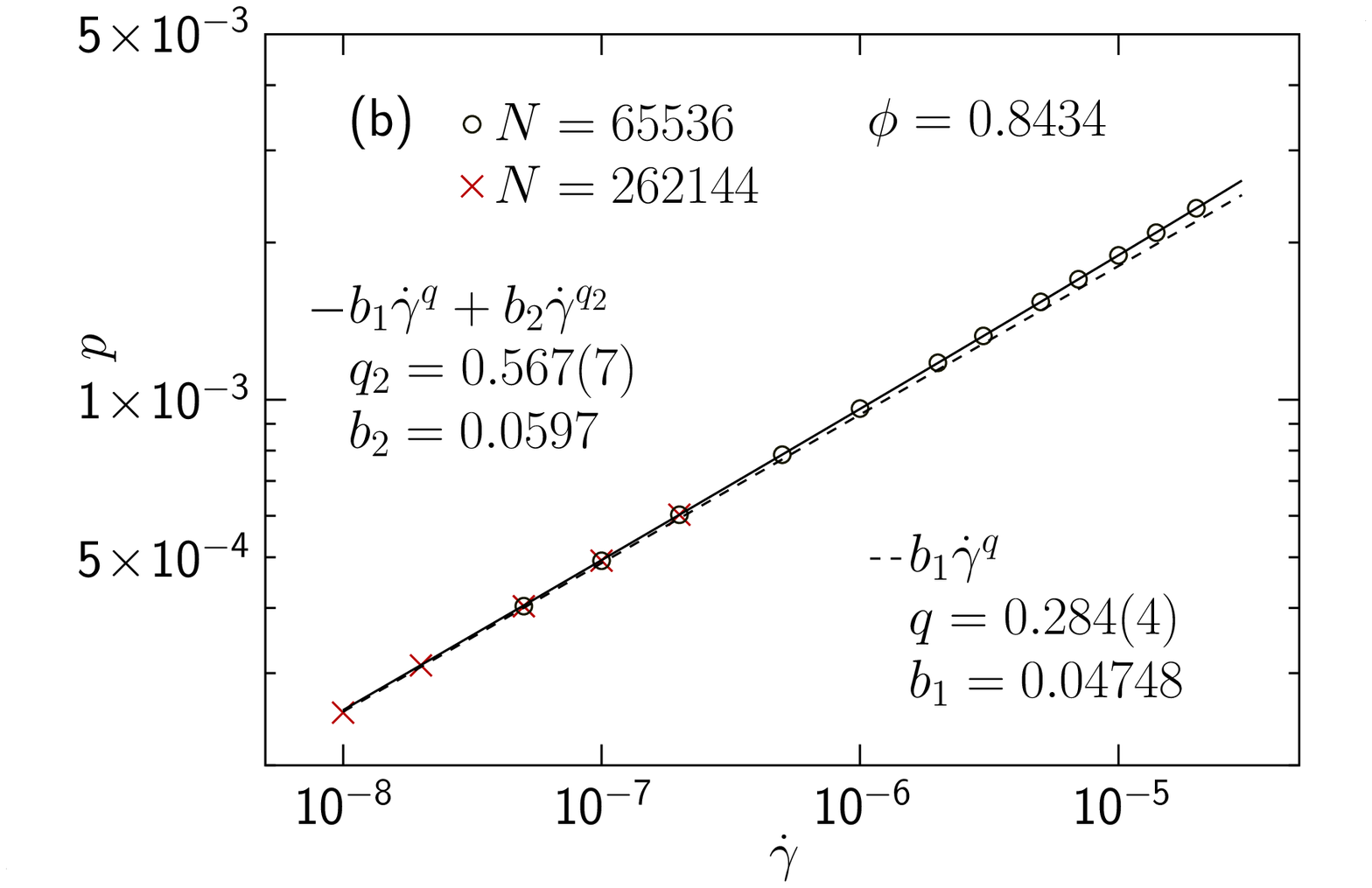}
  \caption{Determination of the exponents $q$ and $q_2$ that characterize the two terms in
    the shear stress. The figures show results from simultaneous fits of $\sigma$ and $p$
    at $\phi=0.8434\approx\phi_J$ to \Eqs{fit-sigma}{fit-p}, demanding that both $q$ and
    $q_2$ are the same, i.e.\ $q_2^{(p)}=q_2$---method C of \App{q_q2}. Panels (a) and (b)
    show $\sigma(\phi_J, \gdot)$ and $p(\phi_J, \gdot)$. The dashed lines are the main
    terms, $\sim\gdot^q$, whereas the solid lines are the full expressions. The simpler
    approach to fit $\sigma(\phi_J,\gdot)$ to \Eq{fit-sigma}, only---this is method A of
    \App{q_q2}---gives just slightly different values of $q$ and $q_2$. Note that the size
    of the secondary terms, in absolute terms, is about the same for both quantities, as
    $b_2\approx a_2$. The \emph{relative} size of the secondary term is however
    considerably smaller for $p$ than for $\sigma$.}
  \label{fig:fit-sigma-gdot-8434}
\end{figure}

The focus of the present paper is not on the values of the exponents and the main
conclusion from \Eq{sigma-scale} is that the shear stress consists of two terms.  In the
analyses below we will take $\phi_J\approx0.8434$ \cite{Heussinger_Barrat:2009,
  Olsson_Teitel:gdot-scale}. We write \Eq{sigma-two} as
\begin{equation}
  \label{eq:fit-sigma}
  \sigma(\phi_J,\gdot) = a_1\gdot^q + a_2\gdot^{q_2} \equiv \sigma_1(\phi_J,\gdot) +
  \sigma_2(\phi_J,\gdot).
\end{equation}
It is now perfectly possible to determine the exponents $q$ and $q_2$ by fitting
$\sigma(\phi_J,\gdot)$ to the middle expression of \Eq{fit-sigma}, but in order to get
higher precision in the determinations we follow \REF{Rahbari_Vollmer_Park} and make use
of the expectation that the same exponents should be present also in the analogous
expression for the pressure,
\begin{equation}
  \label{eq:fit-p}
  p(\phi_J, \gdot) = b_1 \gdot^q + b_2 \gdot^{q^{(p)}_2}.
\end{equation}
The simultaneous fits of $\sigma(\phi_J,\gdot)$ and $p(\phi_J,\gdot)$ with this approach,
when taking $q^{(p)}_2=q_2$, are shown in \Fig{fit-sigma-gdot-8434}, and gives the exponents
\begin{eqnarray*}
  q & = & 0.284(4),\\
  q_2 & = & 0.567(7).
\end{eqnarray*}
The error estimates correspond to three standard deviations.
More details on this approach and some similar methods are given in \App{q_q2}

\begin{figure}
  \includegraphics[width=7cm]{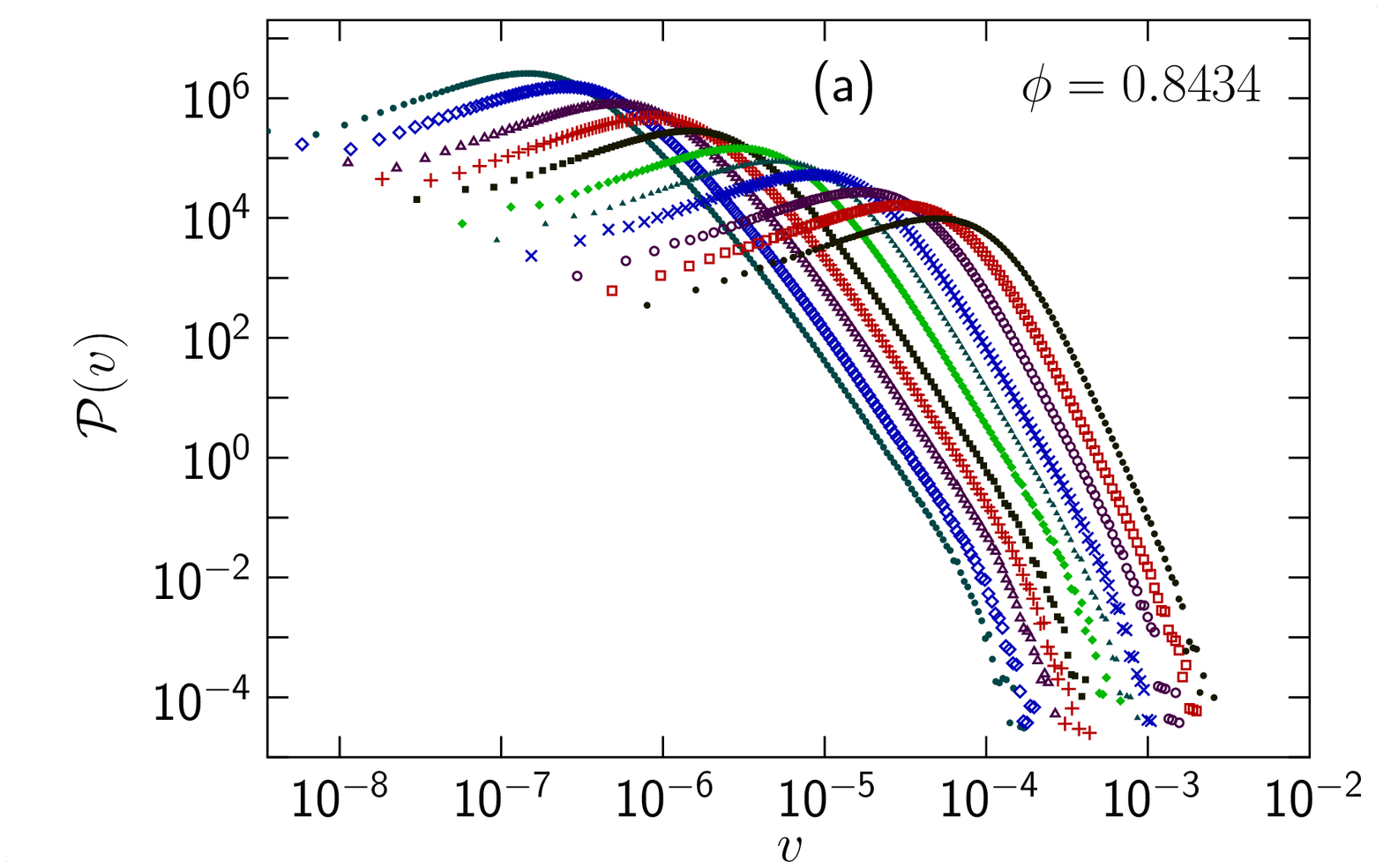}
  \includegraphics[width=7cm]{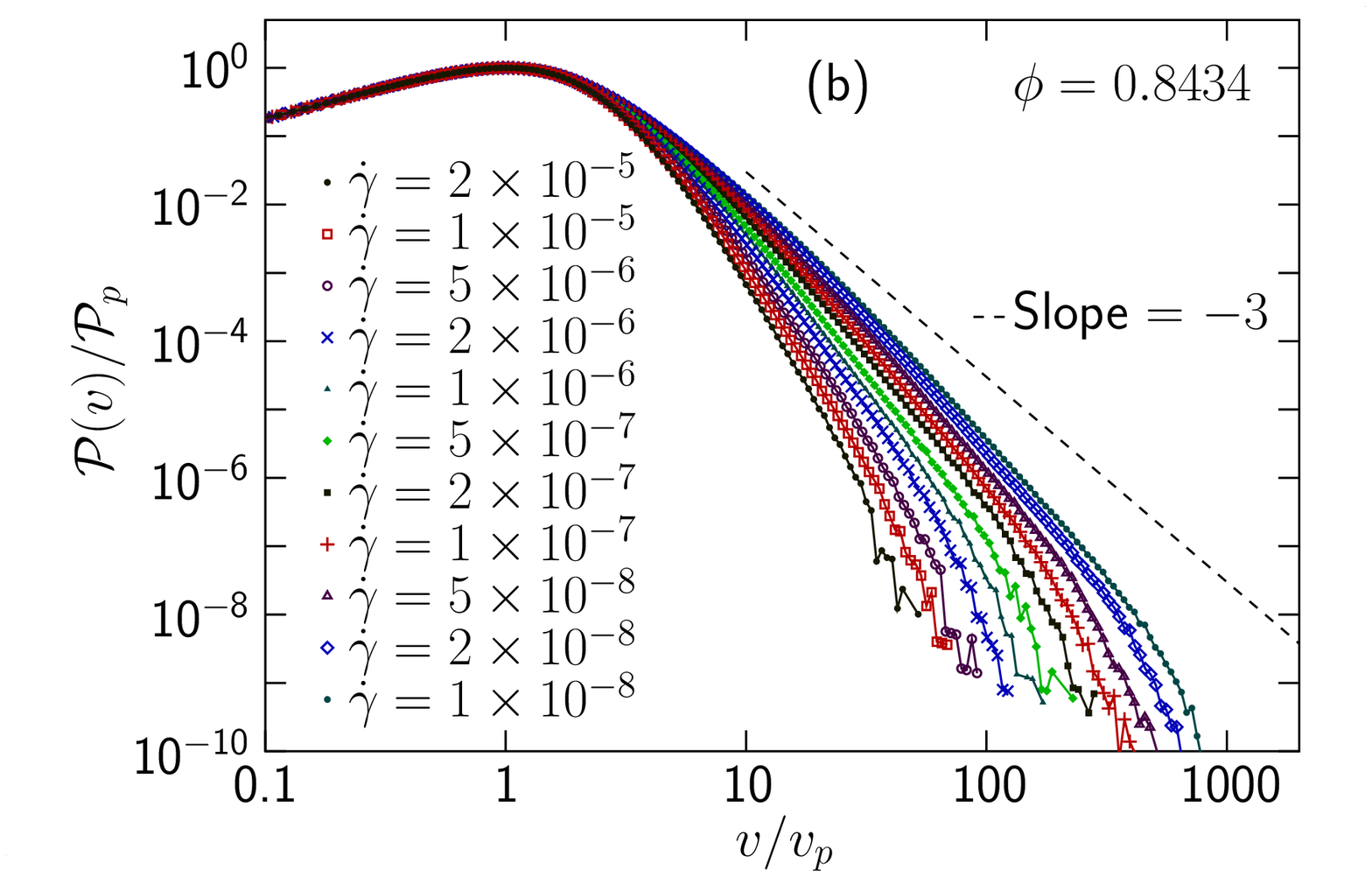}
  \includegraphics[bb=41 322 354 654, width=4.2cm]{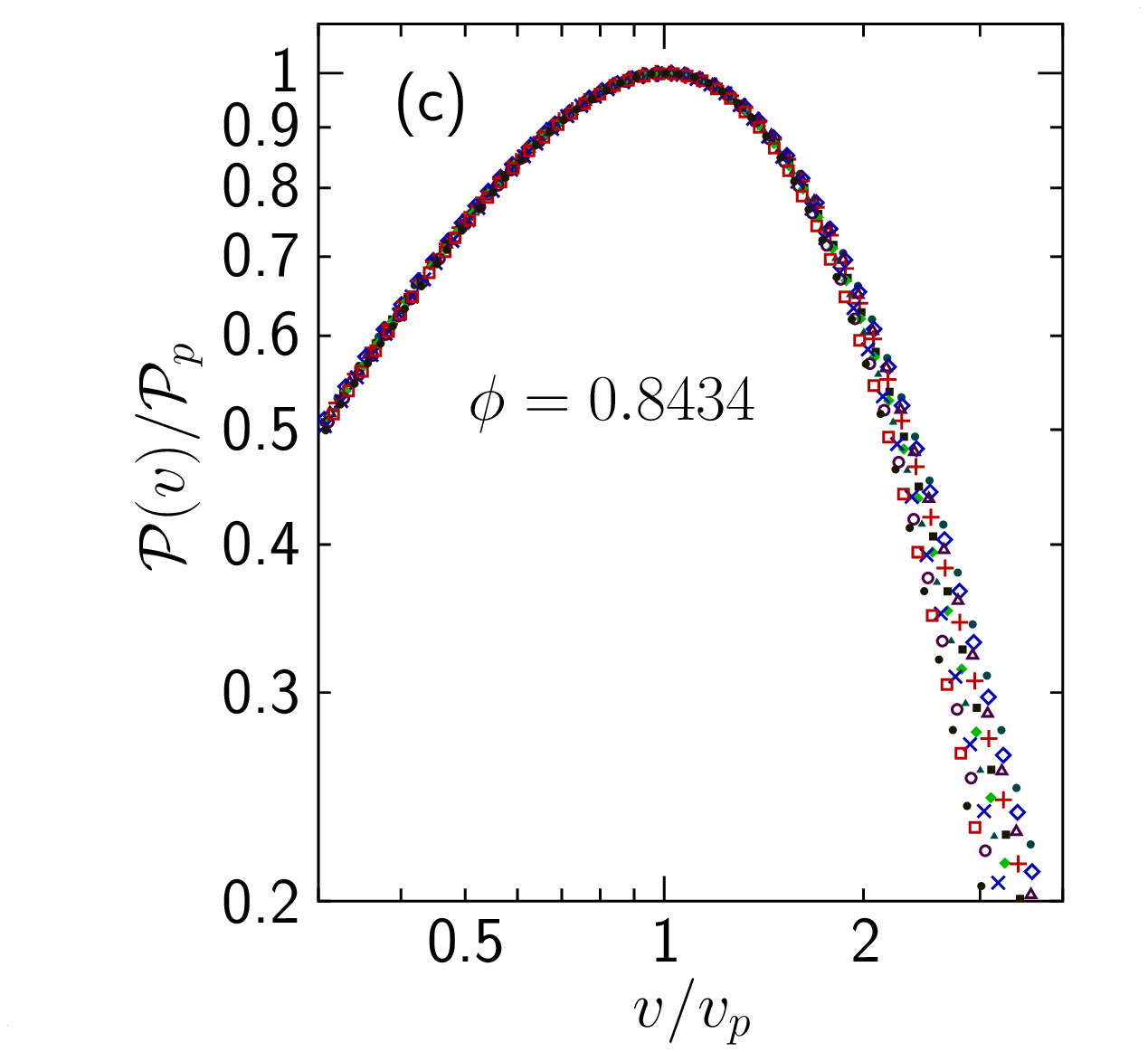}
  \includegraphics[bb=41 322 354 654, width=4.2cm]{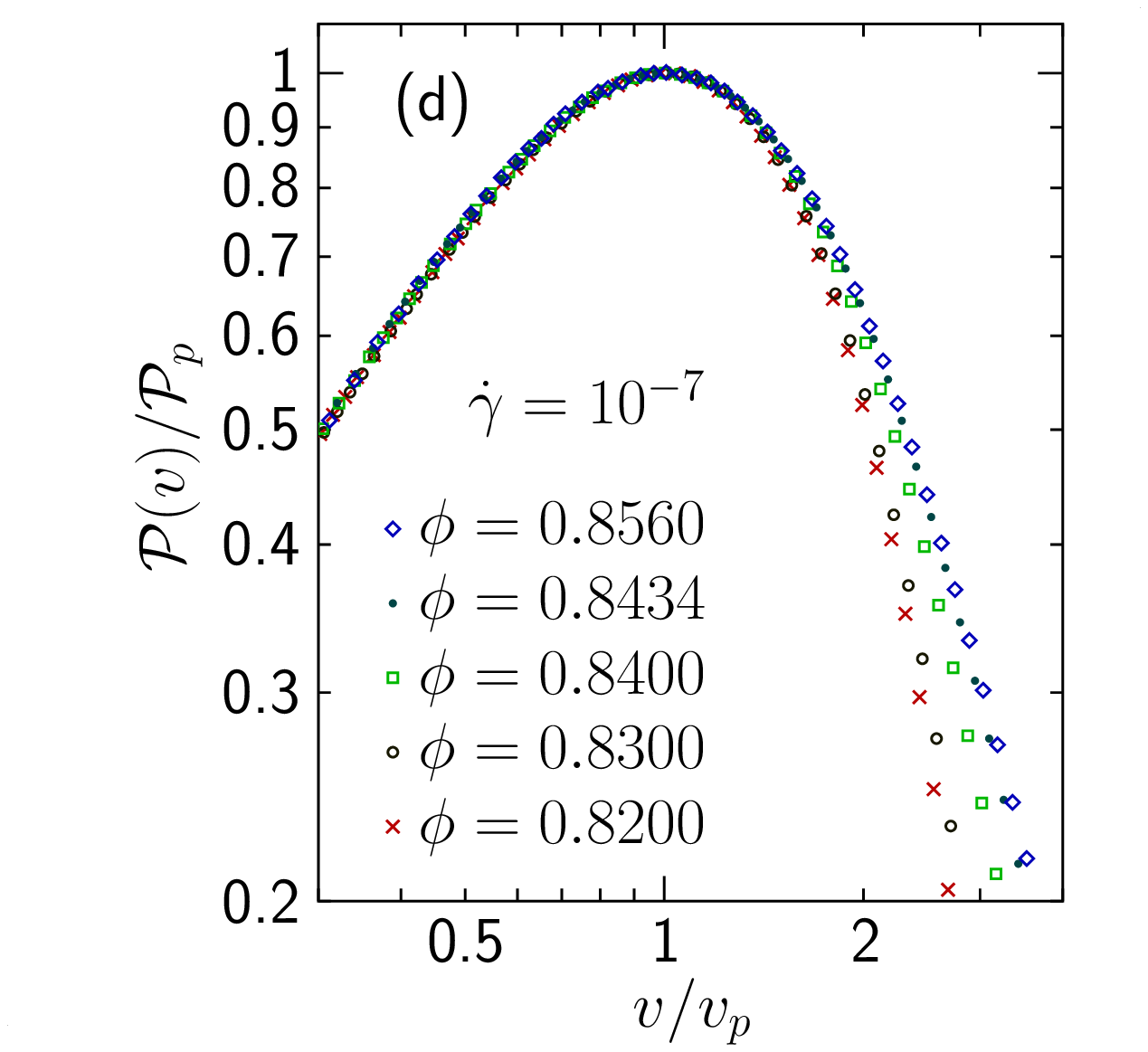}
  \caption{Velocity distribution at $\phi=0.8434\approx\phi_J$ and several different shear
    strain rates. Panel (a) gives $\cP(v)$ for several different shear strain rates.  As
    is clear from panel (a) each data set has a clear peak and panel (b) shows the same
    data rescaled to make the peaks coincide. It is then found that the rescaled $\cP(v)$
    collapse below and up to the peak whereas the data above the peak depend strongly on
    $\gdot$. Panel (c) is a zoom-in on the data of panel (b). Panel (d) shows that the
    same kind of collapse is found also for $\cP(v)$ at $\gdot=10^{-7}$ for $\phi$ both
    below and above $\phi_J$.}
  \label{fig:vhist-v}
\end{figure}

\subsection{Scaling of the peak properties}
\label{sec:peak-exponent}

The velocity distributions at $\phi=0.8434\approx\phi_J$ and for a range of different
shear strain rates from $\gdot=1\times10^{-8}$ through $2\times10^{-5}$ are shown in
\Fig{vhist-v}(a). [Since these figures with double-log scale are not immediately amenable
for simple interpretation, \App{linlog} shows both $\cP(v)$ and a few other quantities on
both logarithmic and linear scales.]  At each $\gdot$ there is a peak in ${\cal P}(v)$ at
low velocities and we identify peak height $\cP_p$ and peak position $v_p$. These
quantities are then used to rescale both axes in the figure such that the peaks fall on
top of each other and, as shown in \Fig{vhist-v}(b) and in the zoomed-in \Fig{vhist-v}(c),
these data collapse nicely up to and slightly above the peak. The same kind of behavior is
found for $\cP(v)$ also at densities away from $\phi_J$ which is clear from
\Fig{vhist-v}(d) which shows the same kind of data for $\gdot=10^{-7}$ and $\phi=0.82$,
0.83, 0.84, 0.8434, and 0.8560. This therefore suggests that the low-velocity part of the
distribution is governed by a simple dynamics with a robust behavior that gives a similar
shape of the distribution independent of detailed properties of the system, as e.g.\
number of contacts. This is in clear contrast to the behavior above the peak where the
distributions are algebraic, $P(v)\sim v^{-r}$, with an exponent that changes with $\gdot$
and $\phi$ and appears to approch $r=3$ at criticality \cite{Olsson:jam-vhist}. (The
distributions are eventually cut off exponentially, which is an effect of the finite
strength of the contact forces that puts a limit on the total net force and thereby on the
velocity \cite{Olsson:jam-vhist}.)

To capture the velocity dependence in the expression for $\sigma$, \Eq{sigma-intv}, we now
introduce $S(v)$ which is the contribution to $\sigma$ from the velocities up to $v$:
\begin{equation}
  \label{eq:S}
  S(v) = \frac{N}{V}\frac{k_d}{\gdot}\int_0^v \cP(v') v'^2 dv'.
\end{equation}
After introducing $x=v/v_p$ and $f(x) = \cP(v)/\cPmax$ the contribution to
$\sigma$ for velocities up to the peak, i.e.\ for all $v<v_p$, becomes
\begin{equation}
  \label{eq:Speak}
  S(v_p) = \frac N V k_d W_p \int_0^1 f(x) x^2 dx,
\end{equation}
which shows that the dependency on $\phi$ and $\gdot$ is only through the peak properties
given by $W_p=\cPmax v_p^3/\gdot$, because the curves for different $\gdot$ and $\phi$
collapse for $v\leq v_p$.

\begin{figure}
  \includegraphics[bb=48 324 320 665, width=4.2cm]{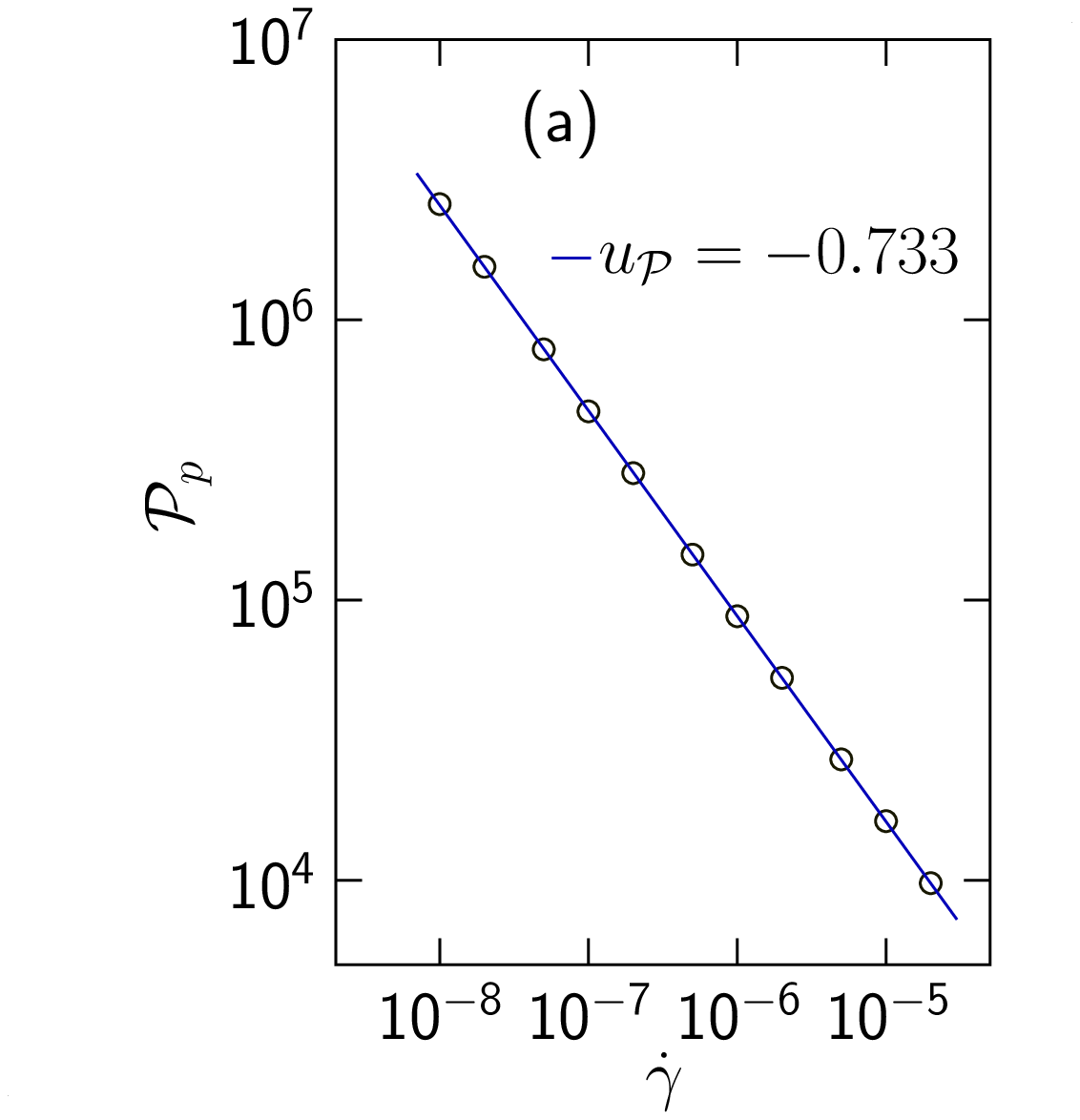}
  \includegraphics[bb=48 324 320 665, width=4.2cm]{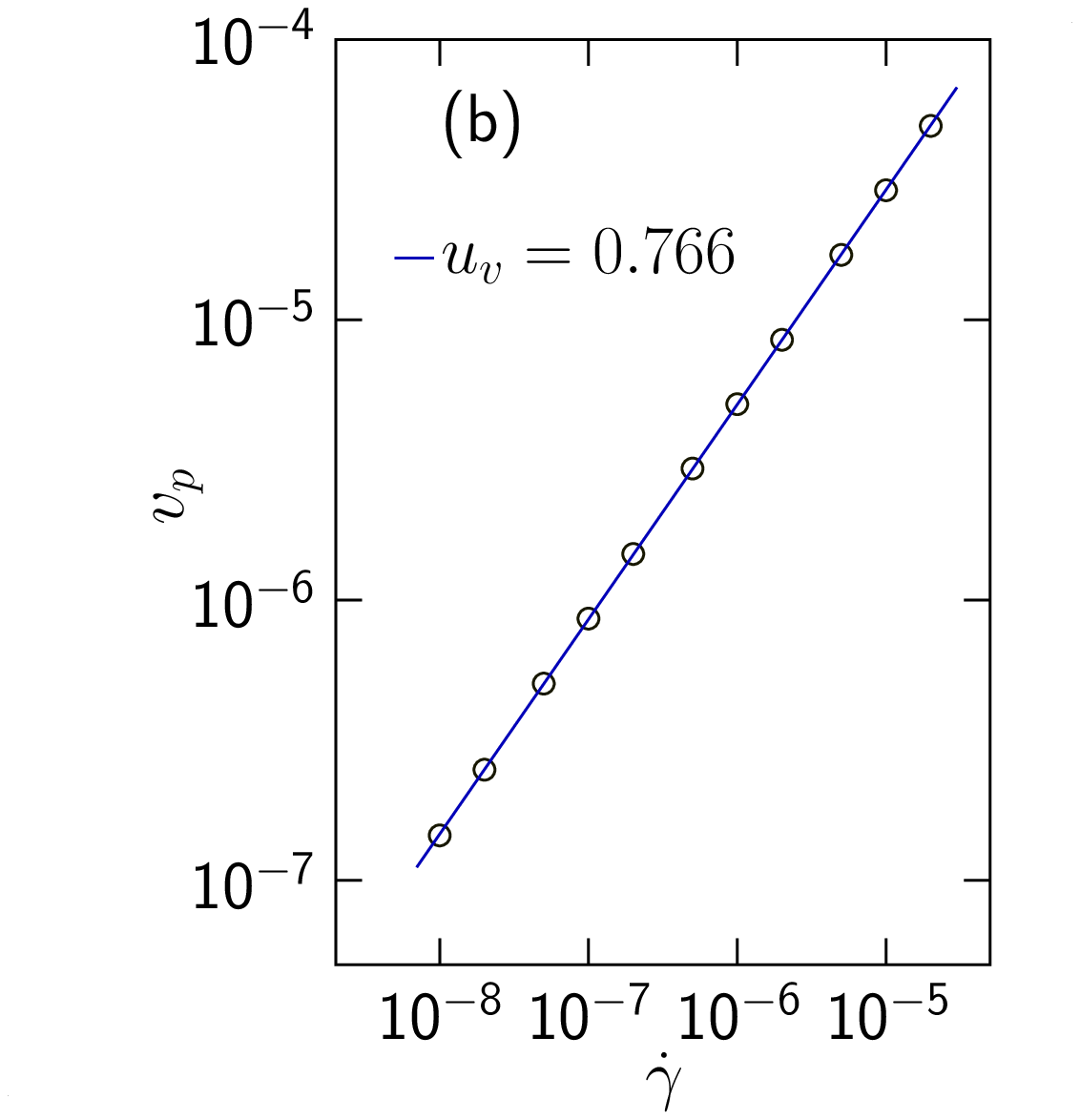}
  \caption{Peak properties at $\phi=0.8434\approx\phi_J$ from \Fig{vhist-v} and
    determinations of the related exponents. Panel (a) is the peak height, $\cP_p$,
    whereas panel (b) is the (velocity) position of the peak, $v_p$.}
  \label{fig:Ppeak_and_vpeak-gdot}
\end{figure}

\Fig{Ppeak_and_vpeak-gdot}, which is again obtained at $\phi=0.8434\approx\phi_J$, shows
that both $\cPmax$ and $v_p$ depend algebraically on $\gdot$ to very good approximations.
We find
\begin{subequations}
  \begin{eqnarray}
    \cPmax(\phi_J,\gdot) & \sim & \gdot^{u_{\cP}}, \quad u_{\cP} = -0.733,
    \label{eq:uP} \\
    v_p(\phi_J,\gdot) & \sim & \gdot^{u_v},\quad u_v=0.766.
    \label{eq:uv}
  \end{eqnarray}
\end{subequations}
For $W_p\equiv \cPmax v_p^3/\gdot$ this gives
\begin{equation}
  \label{eq:Wp}
  W_p(\phi_J,\gdot) \sim \gdot^{u_{\cP}}\gdot^{3u_v}\gdot^{-1} \sim \gdot^{u_w},
\end{equation}
with
\begin{equation}
  \label{eq:uw}
  u_w \equiv 3u_v + u_\cP -1 = 0.565,
\end{equation}
which is in very good agreement with $q_2\approx 0.567$ from the fit of
$\sigma(\phi_J,\gdot)$ to \Eq{fit-sigma}. This therefore suggests that the secondary term,
$\sigma_2$, is related to the slow particles in the peak of the distribution.

\subsection{Magnitude of $\sigma_s$}
\label{sec:peak-magnitude}

\begin{figure}
  \includegraphics[bb=41 322 324 654, width=4.2cm]{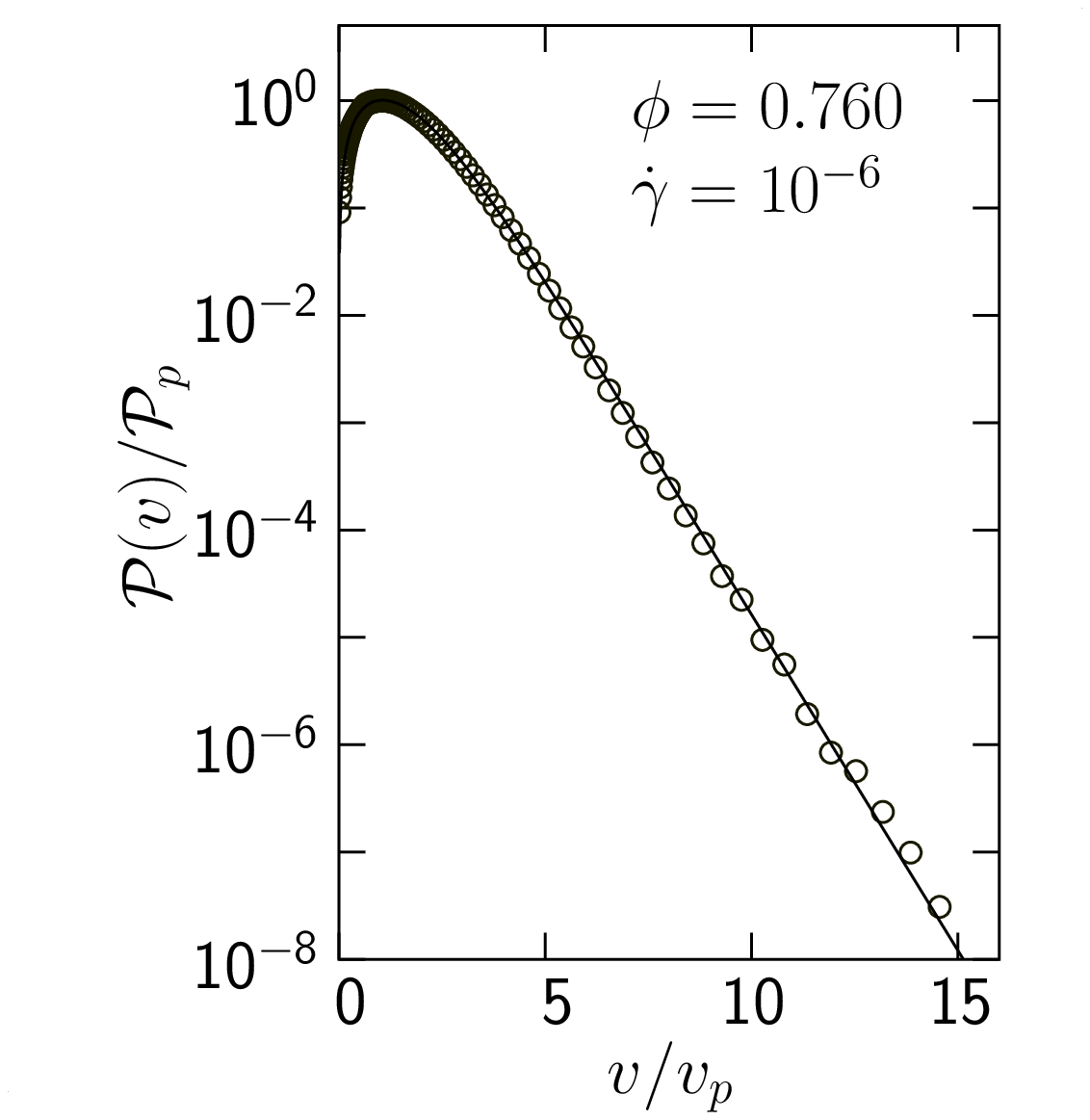}
  \includegraphics[bb=41 322 324 654, width=4.2cm]{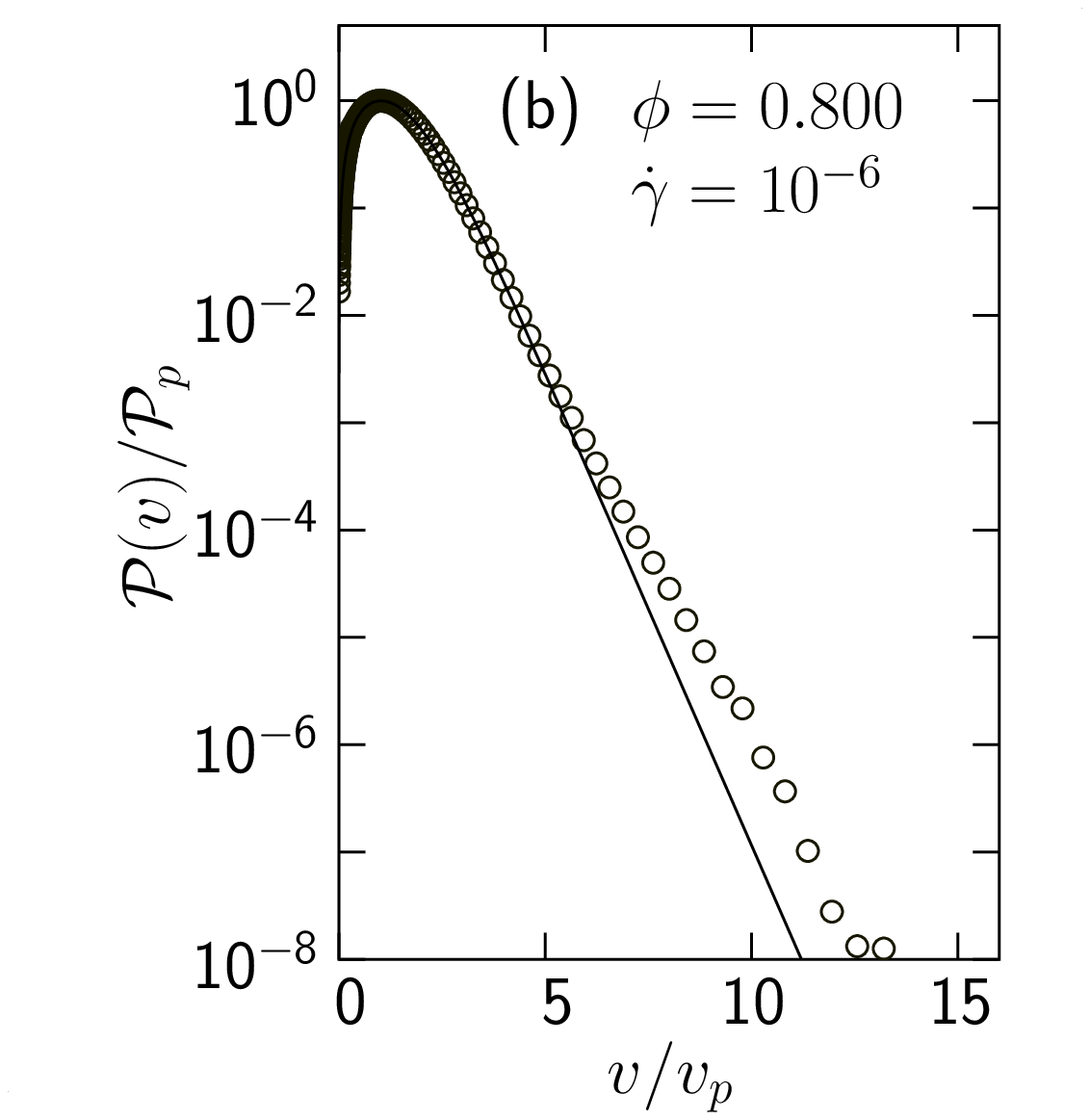}
  \includegraphics[bb=41 322 324 654, width=4.2cm]{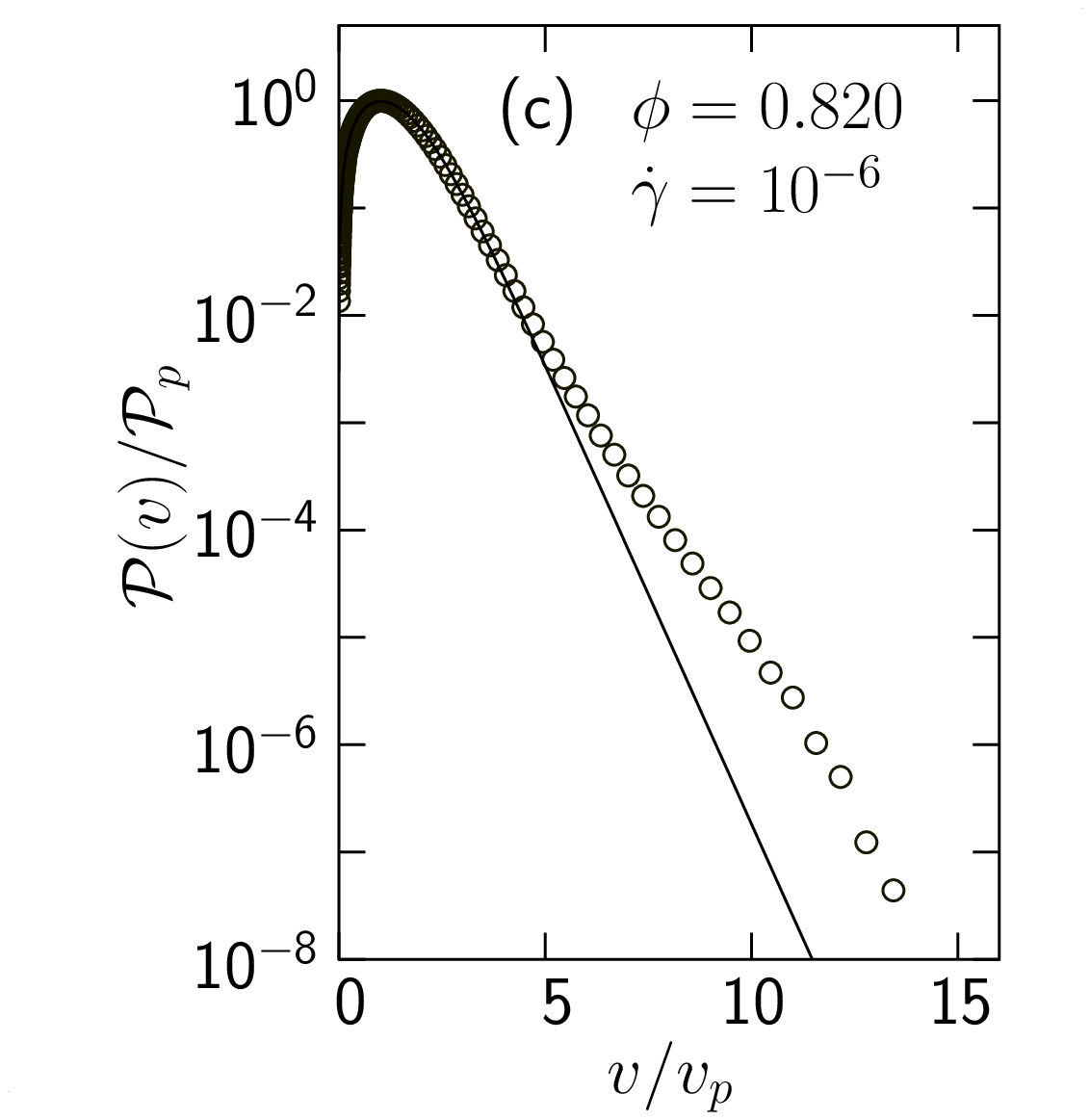}
  \caption{Rescaled velocity distributions at low densities, $\phi=0.76$, 0.80, and 0.82.
    Note that the $x$ axes have linear scales, in contrast to the logarithmic scales in
    \Fig{vhist-v}. At the lowest density, $\phi=0.76$, in panel (a), the distribution is
    exponential whereas there start to develop deviations from that behavior at the higher
    densities in panels (b) and (c). We gather that the exponential decay is the
    characteristics of the slow process whereas the deviations from that behavior develop
    into the algebraic tails of \Fig{vhist-v} that characterize the fast process.}
  \label{fig:P-v-7600-8000-8200}
\end{figure}

We now split the velocity distribution into two terms for the two different processes,
dominated by slow and fast particles, respectively,
\begin{equation}
  \label{eq:Psplit}
  \cP(v) = \cP_s(v) + \cP_f(v),
\end{equation}
where we take $\cP_s(v)=\cP(v)$, for $v\leq v_p$. To get a clue to the shape of $\cP_s(v)$
above the peak, we turn to \Fig{P-v-7600-8000-8200} which shows the velocity distribution
at lower densities, $\phi=0.76$, 0.80, and 0.82. It is there found that the high-velocity
tail shrinks away as $\phi$ is lowered and apparently vanishes at $\phi=0.76$, shown in
\Fig{P-v-7600-8000-8200}(a). What remains is an exponentially decaying $\cP(v)$ and we
take this as a guidance for constructing $\cP_s(v)$ above the peak at general $\phi$.

Defining $\sigma_s$ to be the contribution to $\sigma$ from $\cP_s(v)$,
\begin{displaymath}
  \sigma_s = \frac N V \frac{k_d}{\gdot}\int \cP_s(v) v^2 dv,
\end{displaymath}
and using the same kind of reasoning as in \Eq{Speak}, we introduce
$f_s(x)=\cP_s(v)/\cP_p$ and find
\begin{equation}
  \sigma_s = \frac N V k_d W_p\int f_s(x) x^2 dx = \frac N V k_d W_p I_2,
  \label{eq:sigmas-Wp}
\end{equation}
where $I_2$ is the integral,
\begin{equation}
  \label{eq:I2int}
  I_2 \equiv \int f_s(x)\; x^2\; dx.
\end{equation}

\begin{figure}
  \includegraphics[width=8cm]{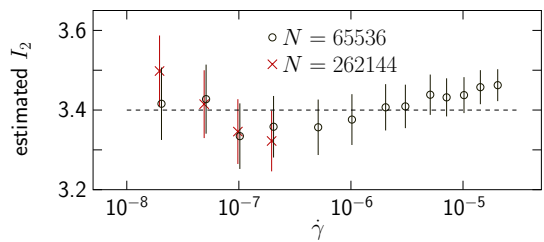}
  \caption{Estimates of $I_2$ from \Eq{I2}. The input for these data are both estimates of
    $\sigma_2(\phi_J,\gdot)\equiv \sigma(\phi_J,\gdot)-a_1\gdot^q$ and $W_p(\phi_J,\gdot)$
    from the velocity distributions. Beside the displayed error bars, which show one
    standard deviation, an important source of error is the uncertainty in $\phi_J$. The
    present estimate is based on assuming $\phi_J=0.843~43$ as obtained in \App{q_q2}.}
  \label{fig:est-I2}
\end{figure}

To determine the numerical value of $I_2$ we assume $\sigma_s=\sigma_2$ and determine
$\sigma_2$ from $\sigma_2=\sigma-a_1\gdot^q$ with $a_1$ and $q$ from the fit to
\Eq{fit-sigma} to get
\begin{equation}
  I_2 = \frac{V}{N}\; \frac{[\sigma(\phi_J,\gdot)-a_1\gdot^q]}{k_d\; W_p(\phi_J,\gdot)},
  \label{eq:I2}
\end{equation}
which is shown in \Fig{est-I2}. Since the size of the secondary term depends sensitively
on the assumed $\phi_J$ we here make use of $\phi_J=0.843~43$ obtained in \App{q_q2}.
Here $\sigma(\phi_J,\gdot)$ from \Eq{sigma} together with $W_p(\phi_J,\gdot)$ from the
peak properties give estimates of $I_2$ for different $\gdot$. We note that the different
estimates of $I_2$ are encouragingly similar and give $I_2\approx3.4$. (The error
bars in \Fig{est-I2} are due to the uncertainties in $a_1$ and $q$ in the fit to
\Eq{fit-sigma}.)

\begin{figure}
  \includegraphics[bb=31 322 364 654, width=4.2cm]{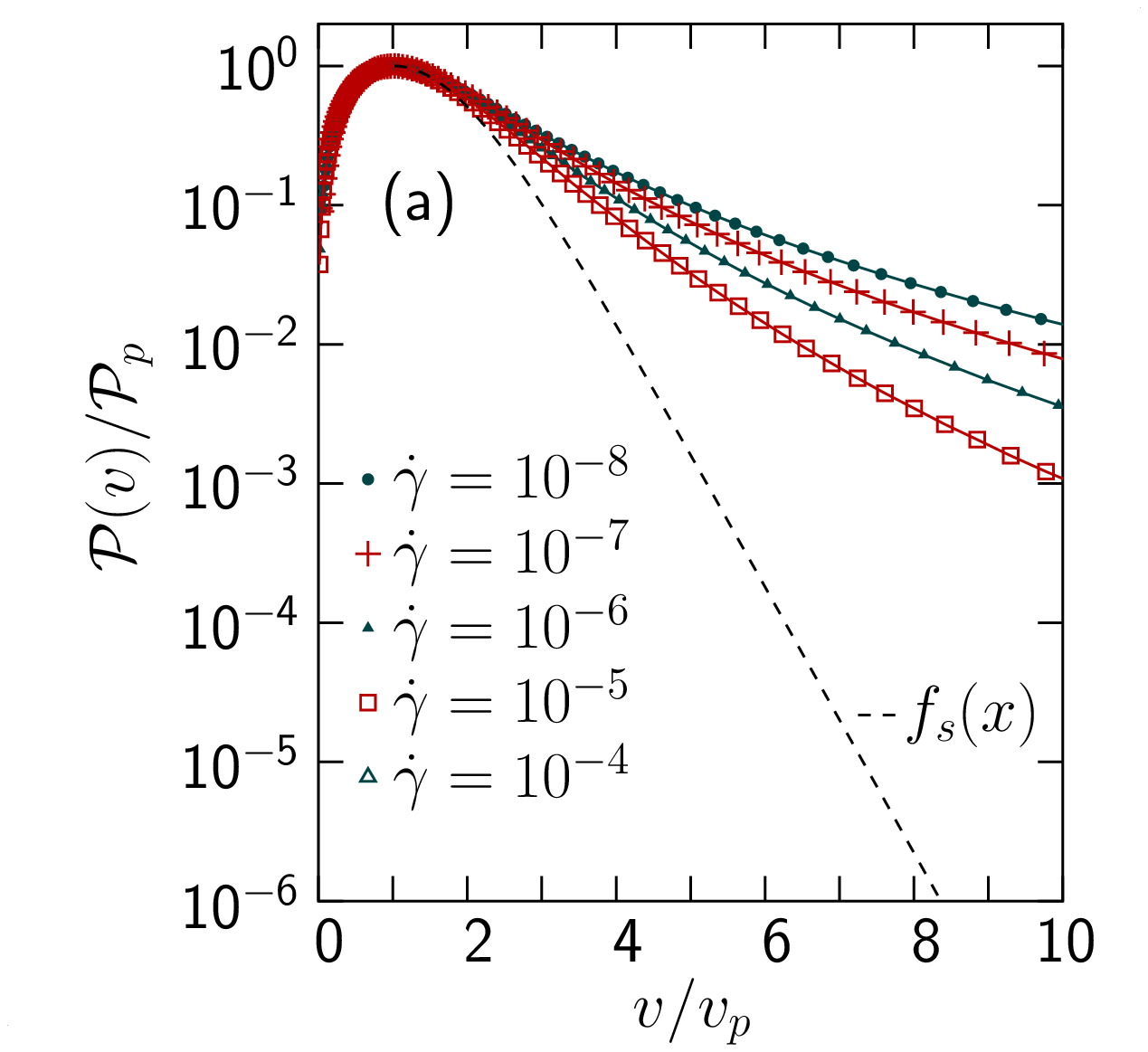}
  \includegraphics[bb=31 322 364 654, width=4.2cm]{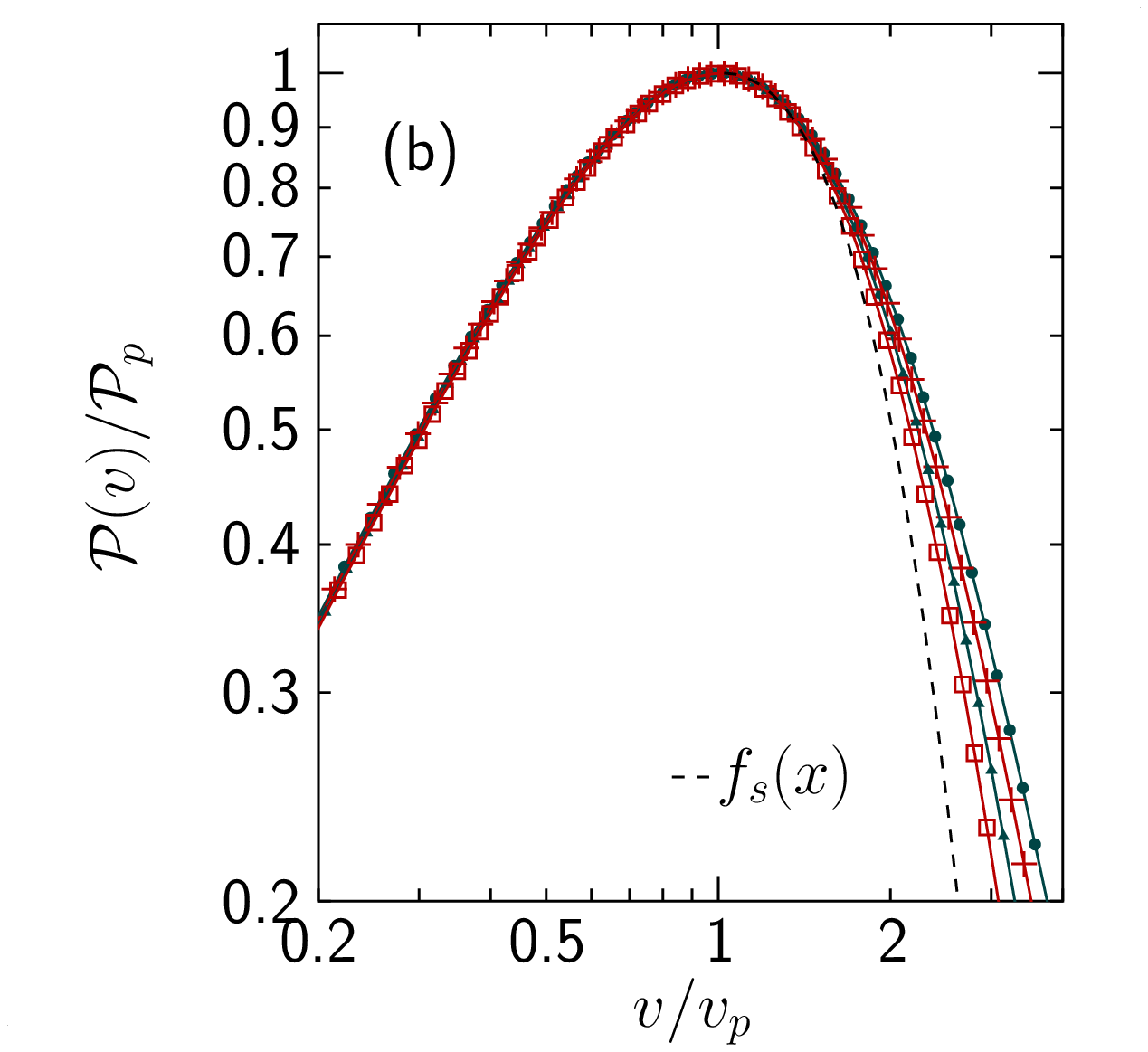}
  \caption{Possible shape of $f_s(x)$ together with data for shear rates $\gdot=10^{-8}$
    through $10^{-4}$. Panel (a) shows the exponential decay of $f_s(x)$ whereas the
    zoom-in in panel (b) shows the same data close to the peak.}
  \label{fig:Pv-v-fs}
\end{figure}

We now take $f_s(x)$ to be given by the rescaled distributions up to (and slightly above)
the peak and assume an exponentially decaying $f_s(x)$ for $x>1$, and adjust the
exponentially decaying part of $f_s(x)$ to give $I_2=3.4$, when integrated with
\Eq{I2int}. The outcome of this procedure is the dashed line in \Fig{Pv-v-fs} which shows
a possible shape of $f_s(x)$.

Before continuing it is worth pointing out that the reasoning above rests on the
assumption that $\cP(v)$ up to the peak is altogether governed by the slow process. Even
though this leads to a consistent picture it should be stressed that there is of course
nothing to preclude the possibility that the distribution for the fast process actually is
small but non-zero at $v=v_p$.

\subsection{Behavior at densities around $\phi_J$}

\begin{figure}
  \includegraphics[width=7cm]{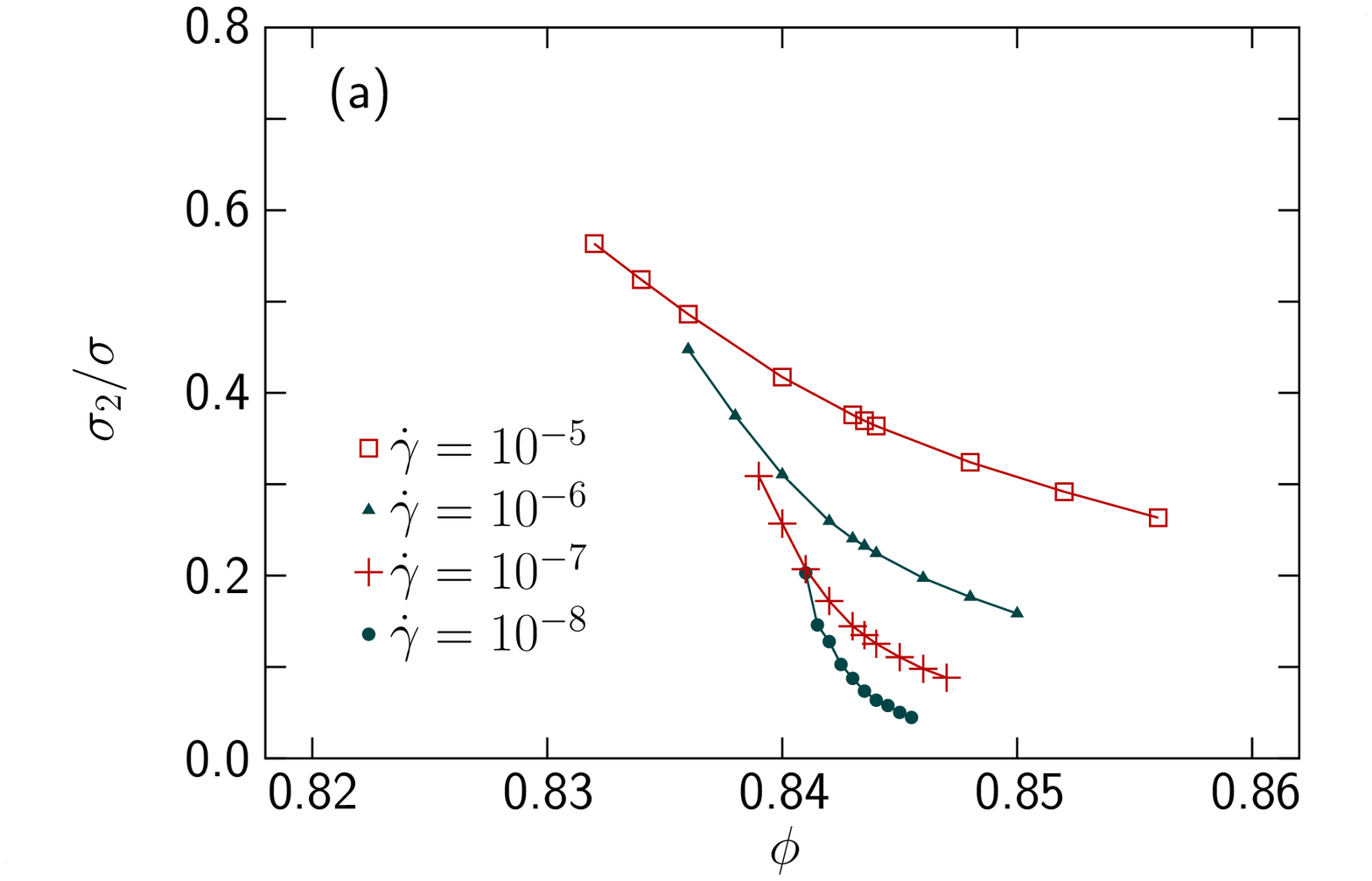}
  \includegraphics[width=7cm]{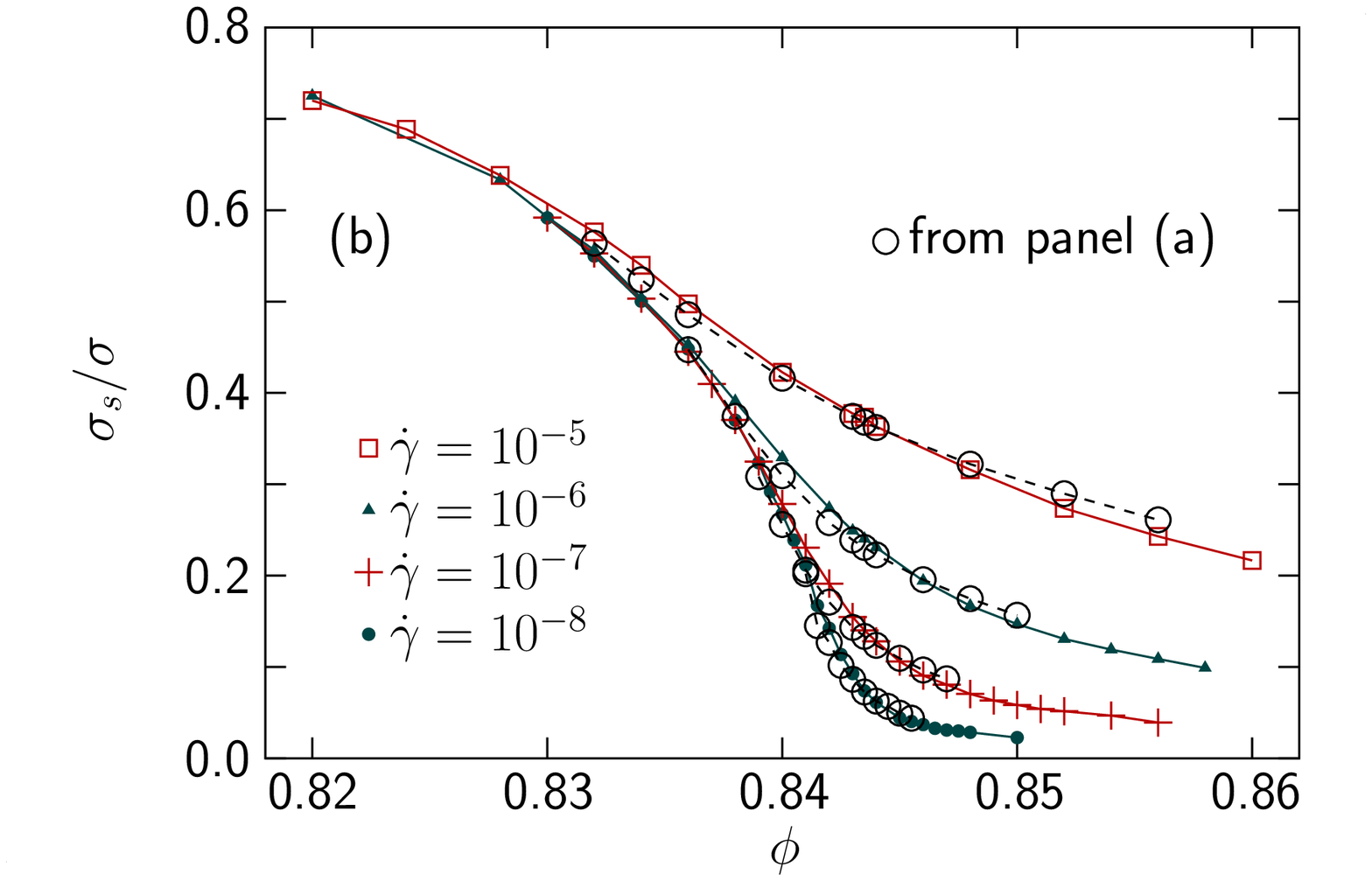}
  \caption{Comparison of $\sigma_2/\sigma$ and $\sigma_s/\sigma$ from two very different
    analyses.  Panel (a) is from the scaling collapse according to \Eq{sigma-scale} where
    $\sigma_2$, as defined in \Eq{fit-sigma}, is the secondary, correction-to-scaling,
    term. Panel (b) is $\sigma_s$ from the peak properties through \Eq{sigmas-Wp} with
    $I_2=3.4$. The open circles connected with a dashed line are the values from panel
    (a). The great similarity of the two quantities suggest that they are related.}
  \label{fig:sigmax.sigma}
\end{figure}

After the analyses of the behavior at $\phi\approx\phi_J$ we now turn to the behavior also
away from $\phi_J$. The aim is not to get reliable determinations of the critical
exponents---such determinations would require both estimates of the uncertainties in $W_p$
and a better understanding of the finite size effects on $\sigma_s$---but rather to show
that $\sigma_s$ from the peak properties through $W_p$ and \Eq{sigmas-Wp} behaves the same
as the secondary term from \Eq{sigma-scale},
\begin{equation}
  \label{eq:sigma2}
  \sigma_2 = \gdot^{q_2} h_\sigma\left(\frac{\phi-\phi_J}{\gdot^{1/z\nu}}\right),
\end{equation}
also away from $\phi_J$.  \Figures{sigmax.sigma}(a) and (b) show the relative
contributions of $\sigma_2$ and $\sigma_s$ and it is clear that they are very
similar. Note that $\sigma_2$---determined from the fit of $\sigma(\phi,\gdot)$ to
\Eq{sigma-scale}---is only available for the range of data that can be used for the fit
whereas $\sigma_s$ can be determined from the peak of the velocity distribution for all
data.

\begin{figure}
  \includegraphics[bb=36 324 354 659, width=4.2cm]{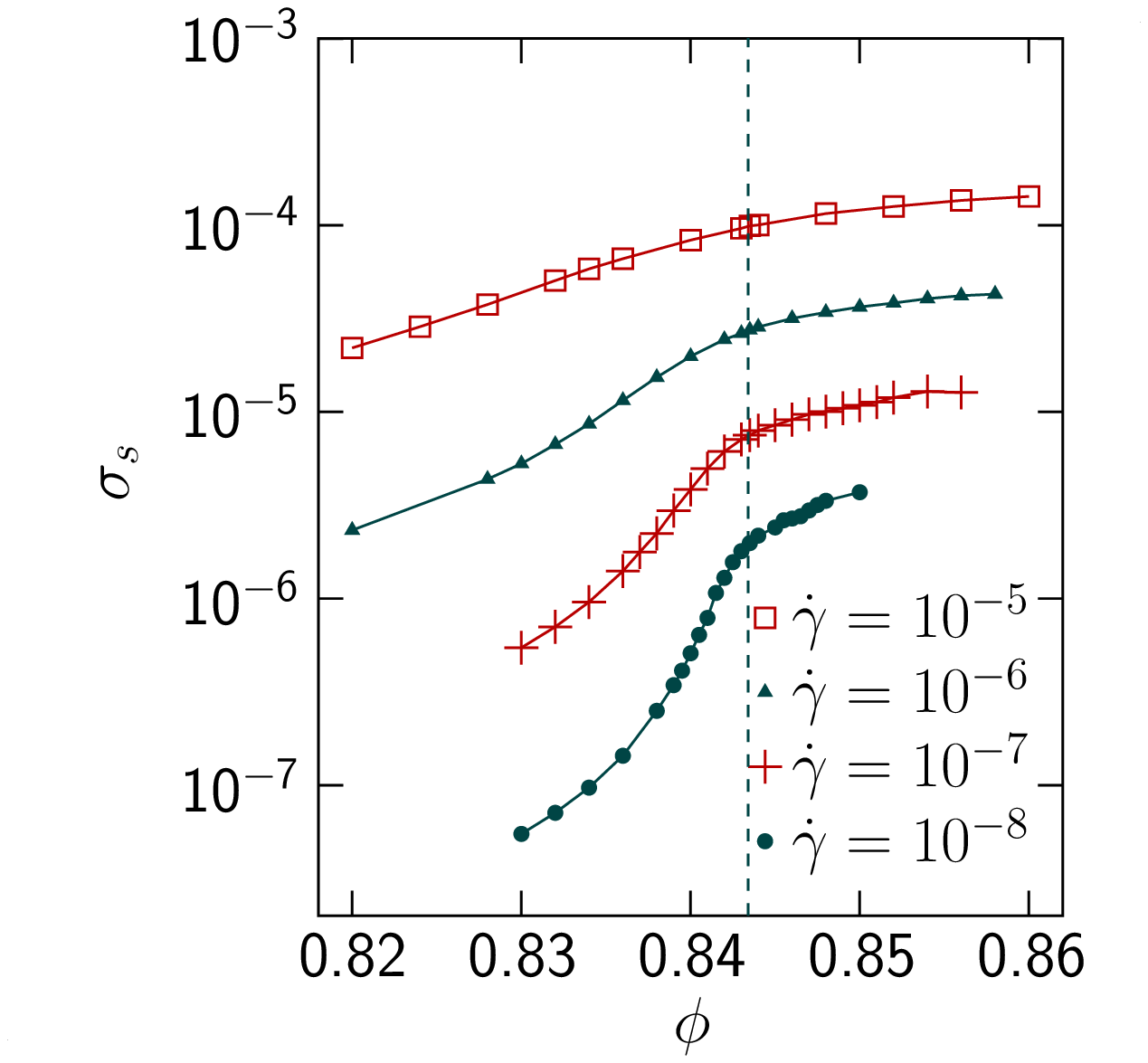}
  \includegraphics[bb=36 324 354 659, width=4.2cm]{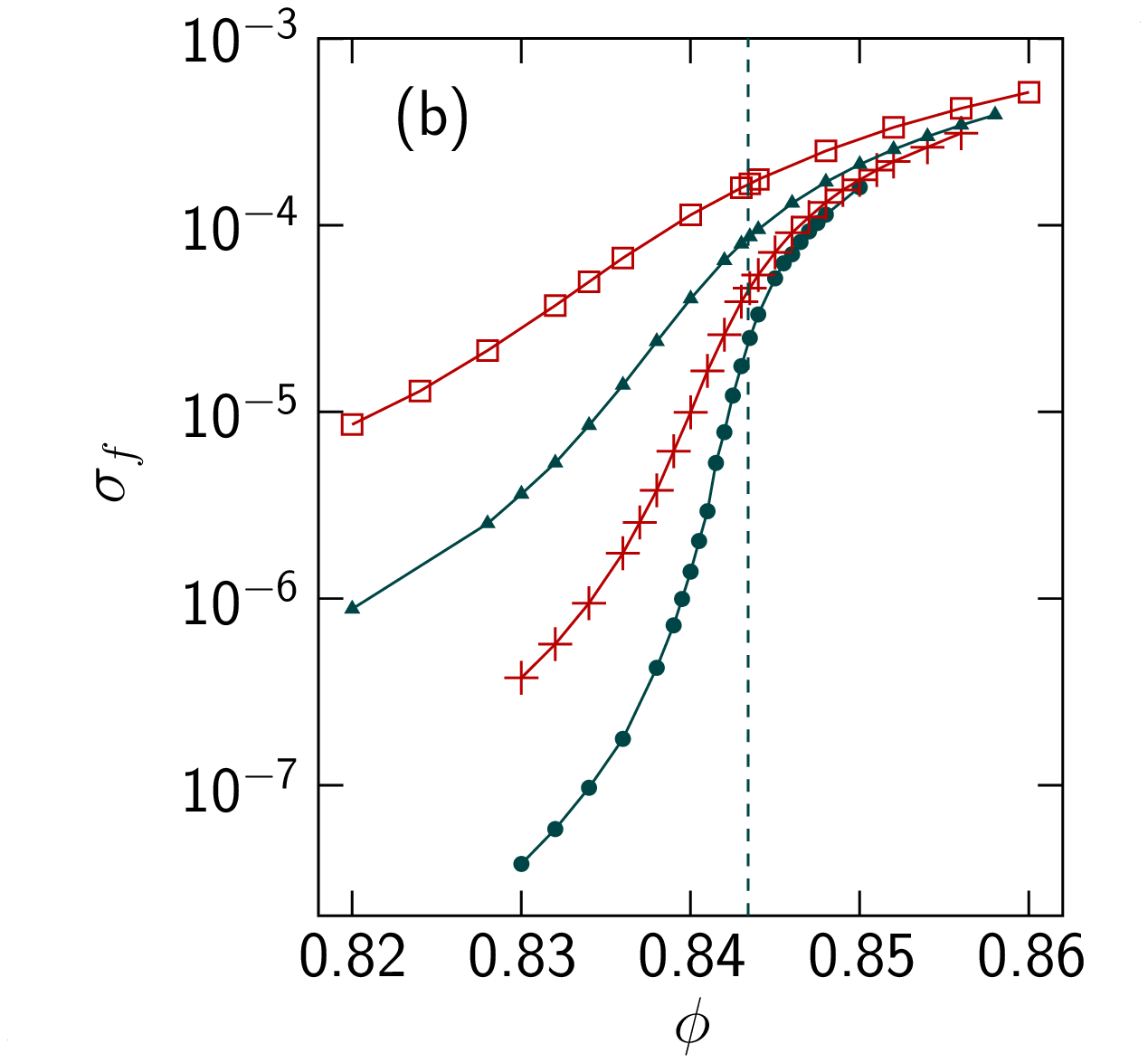}
  \includegraphics[bb=36 324 354 659, width=4.2cm]{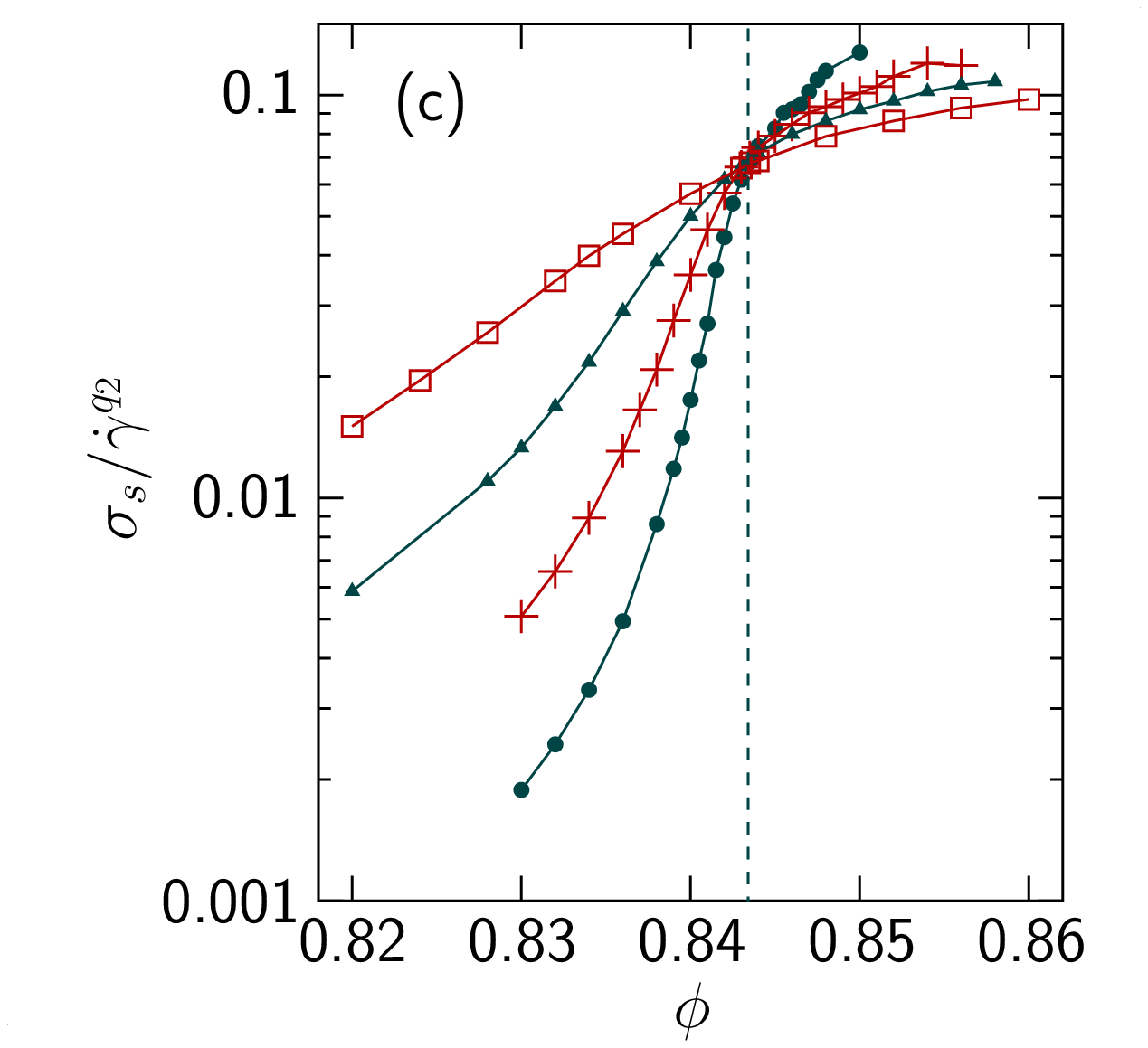}
  \includegraphics[bb=36 324 354 659, width=4.2cm]{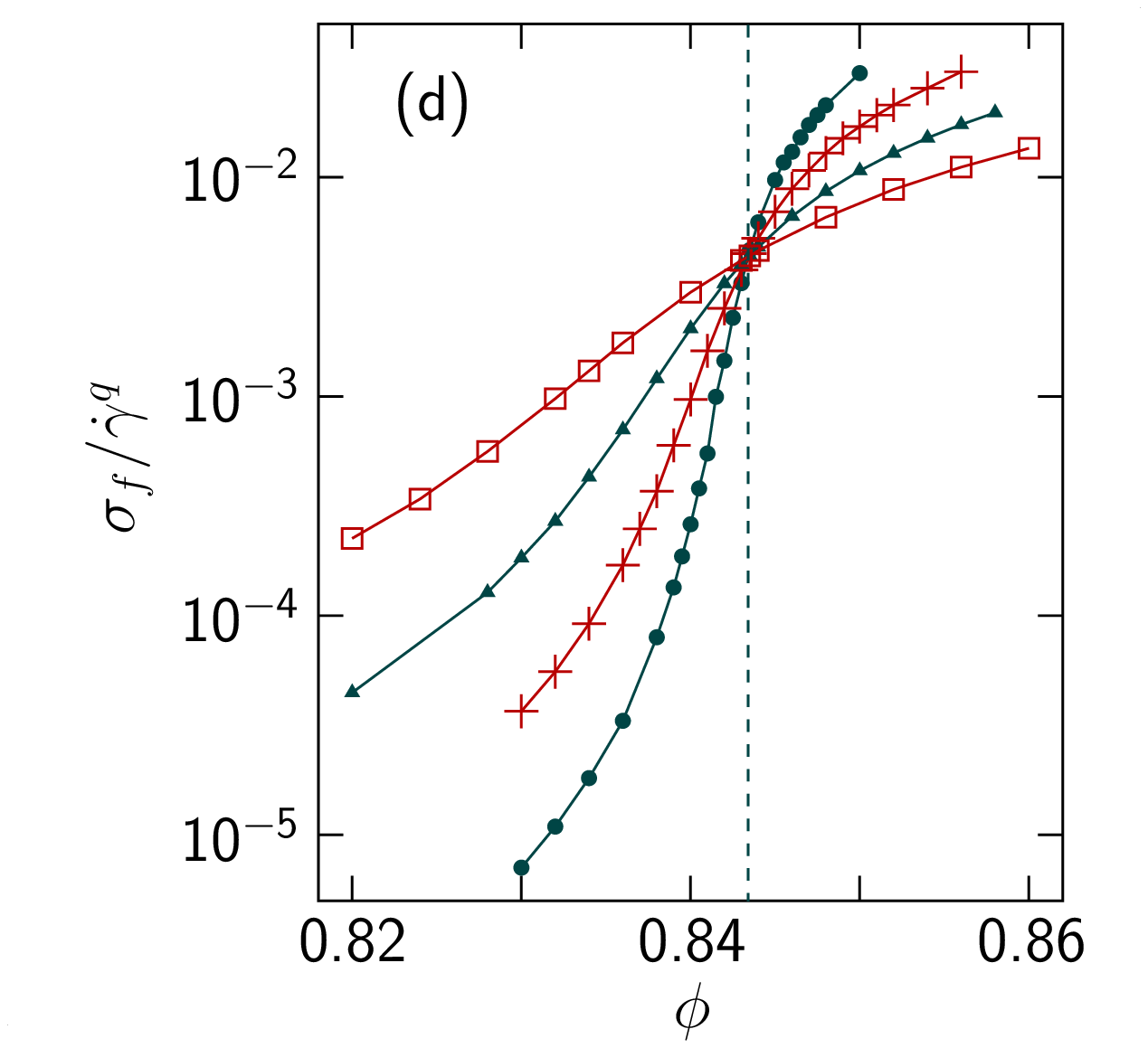}
  \includegraphics[bb=36 324 354 659, width=4.2cm]{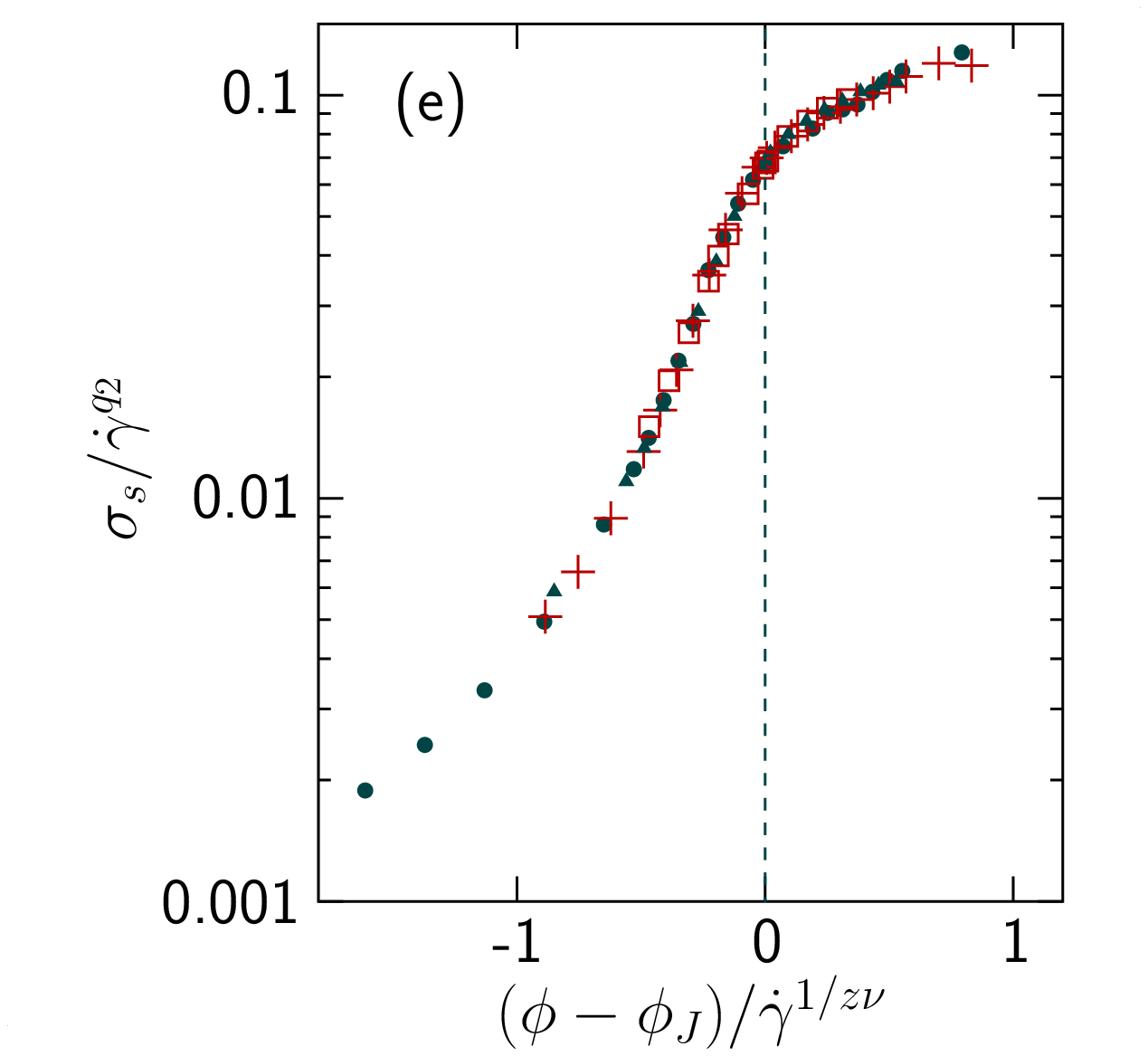}
  \includegraphics[bb=36 324 354 659, width=4.2cm]{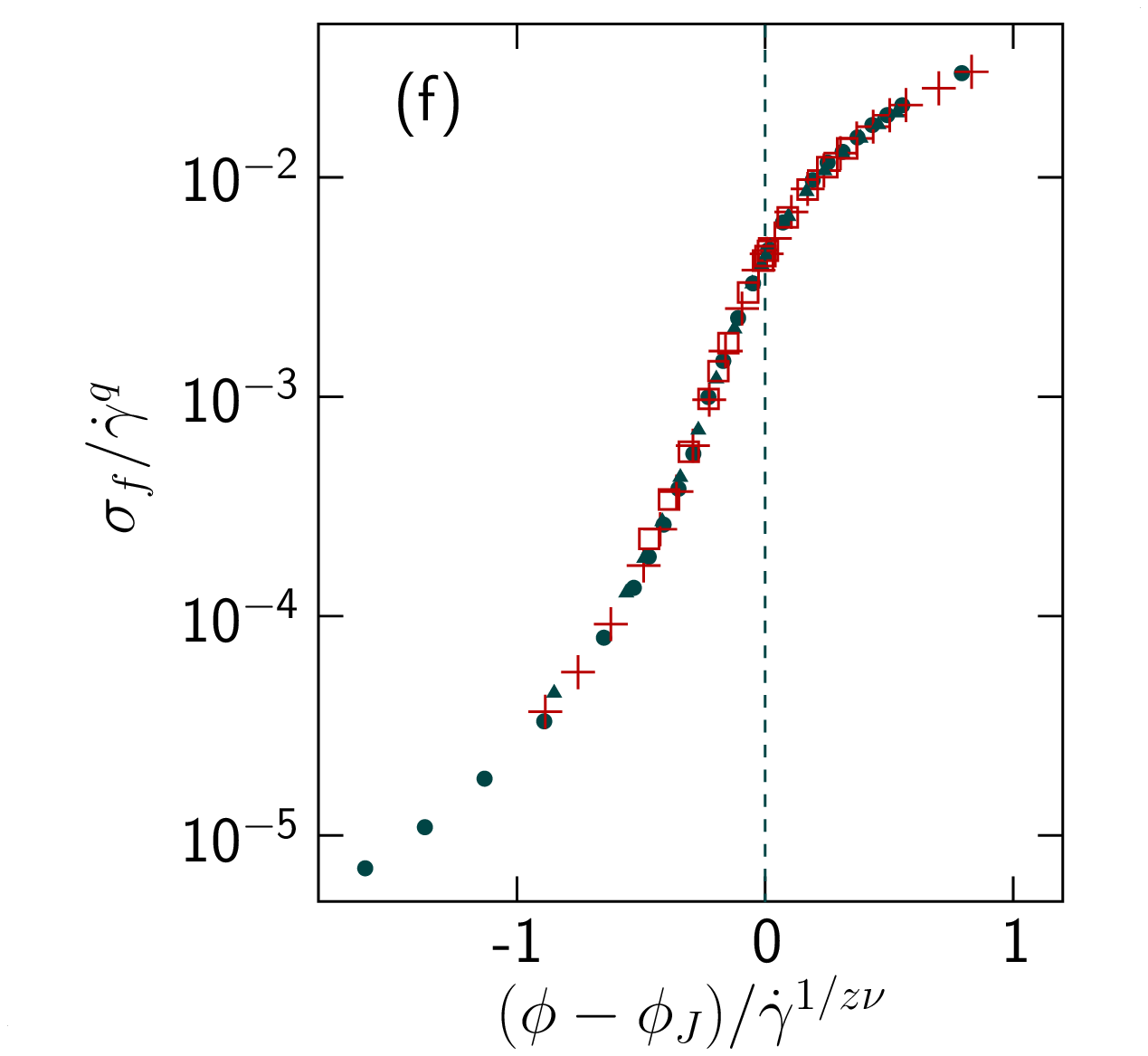}
  \caption{Raw data and scaling analyses of $\sigma_s$ and
    $\sigma_f\equiv\sigma-\sigma_s$. The vertical dashed lines are
    $\phi=0.8434\approx\phi_J$. For clarity we show data for the four shear strain rates,
    only: $\gdot=10^{-8}$, $10^{-7}$, $10^{-6}$, and $10^{-5}$. Panels (a) and (b) are the
    raw $\sigma_s$ vs $\phi$ and $\sigma_f$ vs $\phi$. Panels (c) and (d) are the same
    quantities but scaled by $\gdot^{q_2}$ and $\gdot^{q}$, respectively, which make the
    data cross at $\phi_J$. Panels (e) and (f) are after also rescaling the $x$ axis to
    make the data collapse. Note that $\sigma_s$ are directly from the peak properties as
    the value of $I_2$ just enters as a trivial rescaling parameter. $\sigma_f$, on the
    other hand, also depends on the value of $I_2$ since it controls the size of the
    amounts subtracted from $\sigma$, as shown in \Eqs{sigmas-Wp}{sigmaf}.}
  \label{fig:sigmafs}
\end{figure}

The identification of $\sigma_2$ with $\sigma_s$ means that we should expect $\sigma_s$ to
scale with the exponent $q_2\equiv q+\omega/z$. We introduce $\sigma_f$ which is the
contribution to $\sigma$ due to the fast process,
\begin{equation}
  \label{eq:sigmaf}
  \sigma_f \equiv \sigma - \sigma_s.
\end{equation}
This quantity should---just as the main term---scale with the exponent $q$. \Fig{sigmafs}
shows $\sigma_s$ and $\sigma_f$ vs $\phi$ for $\gdot=10^{-8}$ through $10^{-5}$. Panels
(a) and (b) are the raw data, panels (c) and (d) are the same data rescaled by
$\gdot^{q_2}$ and $\gdot^q$, and panels (e) and (f) show the attempted data collapses when
plotted vs $(\phi-\phi_J)/\gdot^{1/z\nu}$ with $\phi_J=0.8434$ and $1/z\nu=0.26$
\cite{Olsson_Teitel:gdot-scale}. The scaling collapses are very good.

Generally speaking the conclusions arrived at in this way match the results from
\REF{Olsson_Teitel:gdot-scale}. One notable point in \REF{Olsson_Teitel:gdot-scale} is
that $q>1/z\nu$ which implies that $\sigma(\phi,\gdot\to0) \sim (\phi-\phi_J)^y$ where
$y=qz\nu>1$. Though more detailed scaling analyses of $\sigma_f$ and $\sigma_s$ will have
to be deferred to a later paper, we can still attempt a determination of $1/z\nu$ from
$\sigma_f(\phi_J,\gdot)$. This is done by noting that
$\sigma_1=\gdot^q g_\sigma((\phi-\phi_J)/\gdot^{1/z\nu})$ from \Eq{sigma-scale} implies
that
\begin{equation}
  \label{eq:dsigma1dphi}
  \left.\frac{d\ln\sigma_1(\phi,\gdot)}{d\phi}\right|_{\phi_J} \sim \gdot^{-1/z\nu}.
\end{equation}
To estimate $1/z\nu$ we take $\sigma_1=\sigma_f$ and determine the above derivative for
different shear strain rates $10^{-8}\leq\gdot\leq2\times10^{-5}$ by fitting $\ln\sigma_f$
to second order polynomials in $\phi-\phi_J$ for data from narrow intervals around
$\phi_J$, $|\phi-\phi_J|/\gdot^{0.26}<0.3$. From the $\gdot$ dependence of the term linear
in $\phi-\phi_J$ we find $1/z\nu\approx 0.263$ and (with $q=0.284$) $y=qz\nu\approx 1.08$,
in agreement with \REF{Olsson_Teitel:gdot-scale}. It should be noted that the present
approach is much more direct than the scaling analysis \cite{Olsson_Teitel:gdot-scale}
that handles the secondary term through a complicated fitting. In the present approach
that term is eliminated through the peak properties $W_p(\phi,\gdot)$ and the single
parameter $I_2$ from \Eq{I2}.

\subsection{Behavior at $\phi<\phi_J$ as $\gdot\to0$}
\label{sec:hard-particle-limit}

The analyses above are for densities where elasto-plastic processes are important such
that the viscosity is highly rate-dependent and it is interesting to also examine the
behavior in the hard particle region where the viscosity is independent of shear strain
rate. This is reached by taking sufficiently small $\gdot$ at $\phi<\phi_J$.  From the
scaling picture one expects the same analysis to apply also for hard particles below
$\phi_J$, and we here explicitely demonstrate that that actually is the case.

\begin{figure}
  \includegraphics[width=7cm]{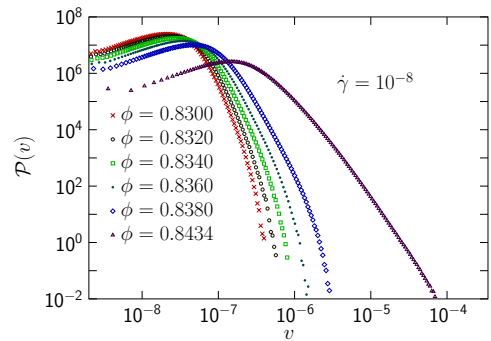}
  \caption{Velocity distributions at the low shear strain rate $\gdot=10^{-8}$ both at
    five densities $\phi=0.830$ through 0.838 representative of the hard disk limit and
    the jamming density, $\phi=0.8434\approx\phi_J$. The properties of the peaks determine
    $W_p=\cP_p v_p^3/\gdot$ which are used in \Eq{sigmas-Wp} with $I_2=3.4$ to estimate
    $\sigma_s$. }
  \label{fig:vhist-8300-8434}
\end{figure}

To approach the hard disk limit we have done simulations of soft disks at densities
$\phi=0.830$ through 0.838 and shear strain rate $\gdot=10^{-8}$ such that the average
overlap of contacting particles is $<10^{-5}d_s$, which means that the simulations are
indeed very close to the hard disk limit. From \Fig{vhist-8300-8434} which is $\cP(v)$
both at five densities $\leq0.838$, well below $\phi_J$, and at $\phi=0.8434\approx\phi_J$
we first note that there is no qualitative difference between the velocity distribution at
$\phi_J$, where the elastic effects are important, and the distribution well below
$\phi_J$, characteristic of the hard disk limit.

\begin{figure}
  \includegraphics[width=7cm]{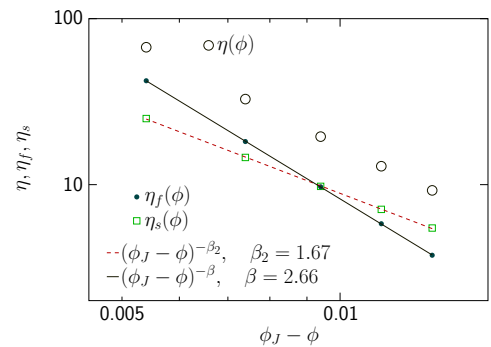}
  \caption{Analyses of the shear viscosity for data in the hard disk limit,
    $\gdot=10^{-8}$ and $\phi=0.830$ through 0.838. The open circles are
    $\eta=\sigma/\gdot$, the open squares are $\eta_s\equiv\sigma_s/\gdot$ with $\sigma_s$
    from the properties of the velocity distributions, as discussed in the caption of
    \Fig{vhist-8300-8434}. The filled circles are $\eta_f\equiv\sigma_f/\gdot$, where
    $\sigma_f=\sigma-\sigma_s$. The fit of $\eta_s$ to an algebraic divergence gives
    $\beta_2=1.67$ whereas the fit of $\eta_f$ gives $\beta=2.66$. As discussed in the
    main text these values are in good agreement with the corresponding values of $q_2$
    and $q$ from the analyses of data at $\phi_J$.}
  \label{fig:eta-83xx}
\end{figure}

\Figure{eta-83xx} shows our results for the viscosity in the hard particle limit. The open
circles are $\eta\equiv\sigma/\gdot$ with $\sigma$ from \Eq{sigma}. The open squares are
$\eta_s\equiv\sigma_s/\gdot$ where $\sigma_s$ is determined with \Eq{sigmas-Wp} with
$W_p=\cP_p v_p^3/\gdot$ from the properties of the peak together with the value
$I_2=3.4$. The solid dots are the contribution from the fast particles
$\eta_f=\eta-\eta_s$. As shown in \Fig{eta-83xx} the values for these exponents from the
fitting of $\eta_s$ and $\eta_f$ below $\phi_J$ to the algebraic divergences (given by the
two terms in \Eq{eta-with-corr}) are $\beta=2.66$ and $\beta_2=1.67$, in very good
agreement with $\beta=2.75$ and $\beta_2=1.67$ from \Eq{beta_beta2}, $1/z\nu=0.26$, and
the values of $q$ and $q_2$ given below \Eq{fit-p}.

The conclusion from the section is thus that the splitting of data into slow and fast
particles works the same for hard particles as for the data around $\phi_J$ and also that
these different determinations of the exponents are in very good agreement.

\subsection{Fast particles}

\begin{figure}
  \includegraphics[bb=70 338 413 660, width=4.2cm]{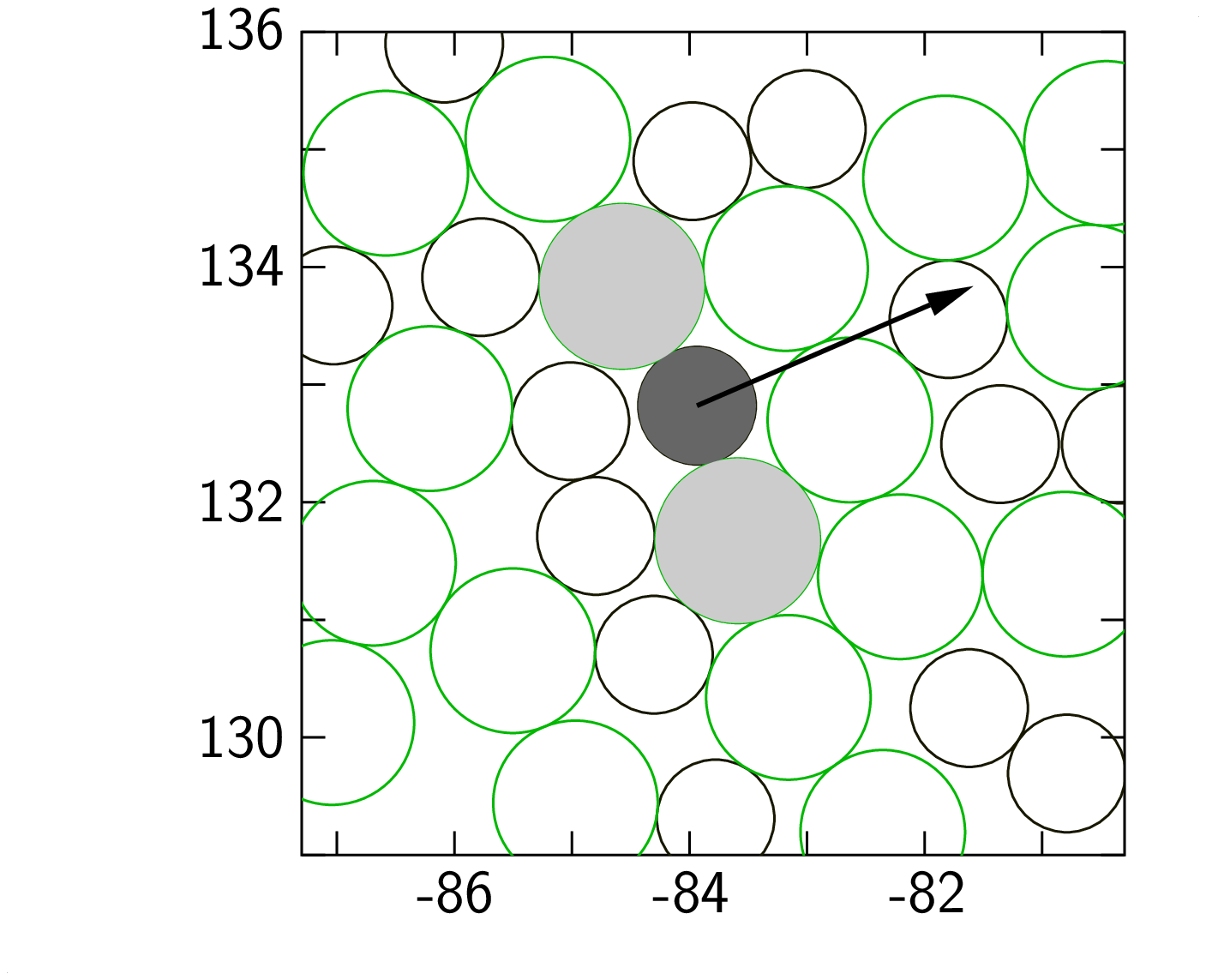}
  \includegraphics[bb=70 338 413 660, width=4.2cm]{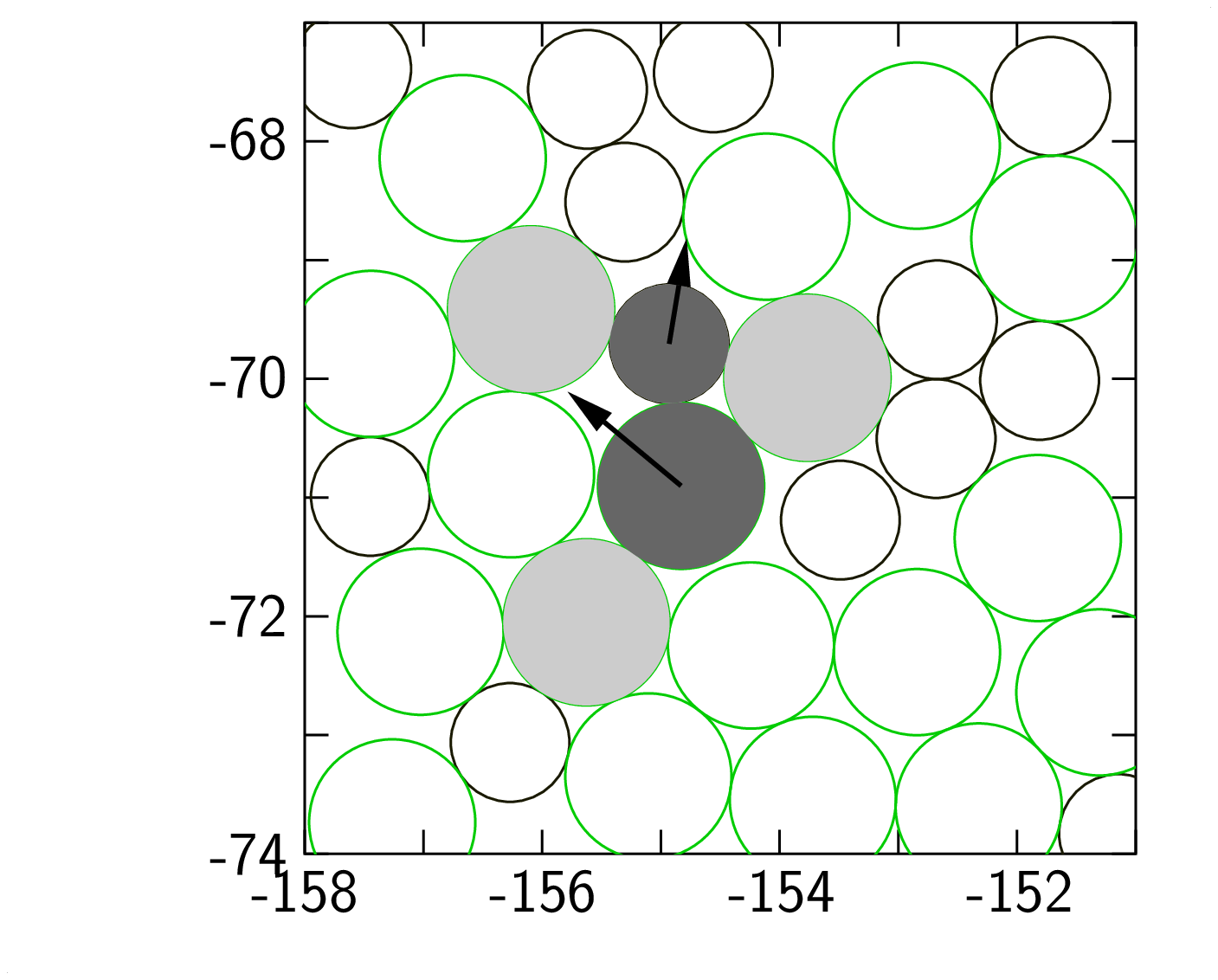}
  \caption{Configuration with fast particles, shown by dark gray. Panel (a) shows a
    particle with velocity $v/\expt v\approx 8.5$. The reason for its high velocity is
    that it is squeezed between the two other particles, shown by light gray, and is
    therefore not in a force-balanced state. Panel (b) shows a configuration with two fast
    particles where a large net force on the big dark gray particle pushes on the small
    dark gray particle, which happens to be free to move and therefore also gets a high
    velocity.}
  \label{fig:snapshot}
\end{figure}

After this comparison of the properties of the peak in the velocity distribution and the
secondary term, as determined from the scaling analysis of $\sigma(\phi_J,\gdot)$ together
with an analysis in the hard disk limit below $\phi_J$, we now turn to the high velocity
regime and the main process, to try to understand the origin of the highest velocities far
out in the tail of the distribution. To that end we have examined several configurations
with fast particles at density $\phi=0.80$. A typical case is as in \Fig{snapshot}(a),
where the fast particle, shown in dark gray, only has two contacting particles and is
therefore in an unbalanced configuration. Since the contact forces in this particular case
are quite large and the three particles are not entirely in line this configuration gives
a large net force on the gray particle and thereby a high velocity. (In \App{wide} we
comment on the understanding that the wide velocity distribution should be related to the
system going back and forth between jammed and unjammed states, and argue that it is not a
tenable explanation.)

Though a single unbalanced particle is the simplest case, the two dark gray particles in
\Fig{snapshot}(b) also have high velocities. In this case a large net force on the big
dark gray particle also makes the small dark gray particle move, and this kind of behavior
may sometimes extend to chains of several particles. It should however be noted that a
bigger number of particles give lower velocities for the same driving force. The tentative
conclusion from this study is thus that the fast process is due to particles being
squeezed, which is in contrast to getting their velocities by being pushed by other
contacting particles with similar velocities.

A consequence of this picture is the presence of an additional time scale, related to the
typical contact force, beside the time scale given by the shear strain rate. This is then
a property which these particles have in common with avalanches that develop according to
their intrinsic dynamics once they are set into motion.

It is interesting to note that two different times scales have previously been found in
analyses of the auto-velocity correlation function \cite{Olsson:jam-vvt}, where one of the
time scales is directly related to the shear strain rate whereas the other is the
``internal time scale'', $t_\mathrm{int} \sim 1/\sigma$. The conclusion that the dynamics
of the fast particles in \Fig{snapshot} is governed by a time scale related to the contact
force, fits well together with $\sigma\sim\expt{f_{ij}}$.

The examples discussed above are for the simple case of the fastest particles far out in
the tail of the distribution, but it is less clear if it is possible to separate all
particles into ``fast'' and ``slow'', as would seem to be required by the splitting of the
velocity distribution into two terms as in \Eq{Psplit}. One attempt in that direction
would be to start from the picture that most particles---the slow ones---move around by
being pushed by other particles with similar velocities and that the squeezing give rise
to ``fast'' particles. One would however also need to characterize a particle as fast if
it is pushed by another fast particle, but it is at present not clear if it is possible to
device reasonable and useful criteria for such splitting into slow and fast
particles. Another possibility would be to give up the idea of a strict splitting of
particles into two disjunct categories, and instead say that any given particle may
participate in, or be affected by, both the fast and the slow process.

\subsection{Spatial velocity correlations}
\label{sec:spatial}

When the correlation length has been identified, one expects that the finite size
dependence should be controlled by the dimensionless ratio $\xi/L$, where $L$ is the
linear system size. In shear-driven jamming this does however not work out as expected.
One example from the literature is in an attempted finite size scaling analysis at
$\phi_J$ \cite{Vagberg_OT:jam-cdrd} where a decent collapse was found when data from
different $L$ were plotted vs $L/\gdot^{-1/z}$, with $z=6.5$, which is clearly different
from the expected $z=1/0.26=3.85$. (As discussed in the jointly published Letter
\cite{jointPRL} this difficulty is resolved by including a correction-to-scaling
term. This finite size scaling does however work differently than commonly expected.)
Another example that is difficult to reconcile with the expected behavior is a recent
examination of the finite size dependence of data in a density range well below $\phi_J$,
where the onset of finite size effects appeared at a constant $L$, even though the
correlation length changes by more than a factor of two across the density interval in
question \cite{Olsson:jam-NIB}.

In critical phenomena one expects a direct link between the diverging correlation length
and the diverging order parameter. As discussed above the shear viscosity is dominated by
the fastest particles and we will now argue that the correlations are instead dominated by
the slower particles, which is thus in contrast to this usual picture. To demonstrate that
the correlations are dominated by slower particles we will use two sets of data, the
``overlap function'' and the velocity correlation function. The former has been widely
used in the literature but the advantage of the latter is that it allows for a more direct
interpretation in terms of the particle displacements.

\begin{figure}
  \includegraphics[bb=41 319 532 545, width=7cm]{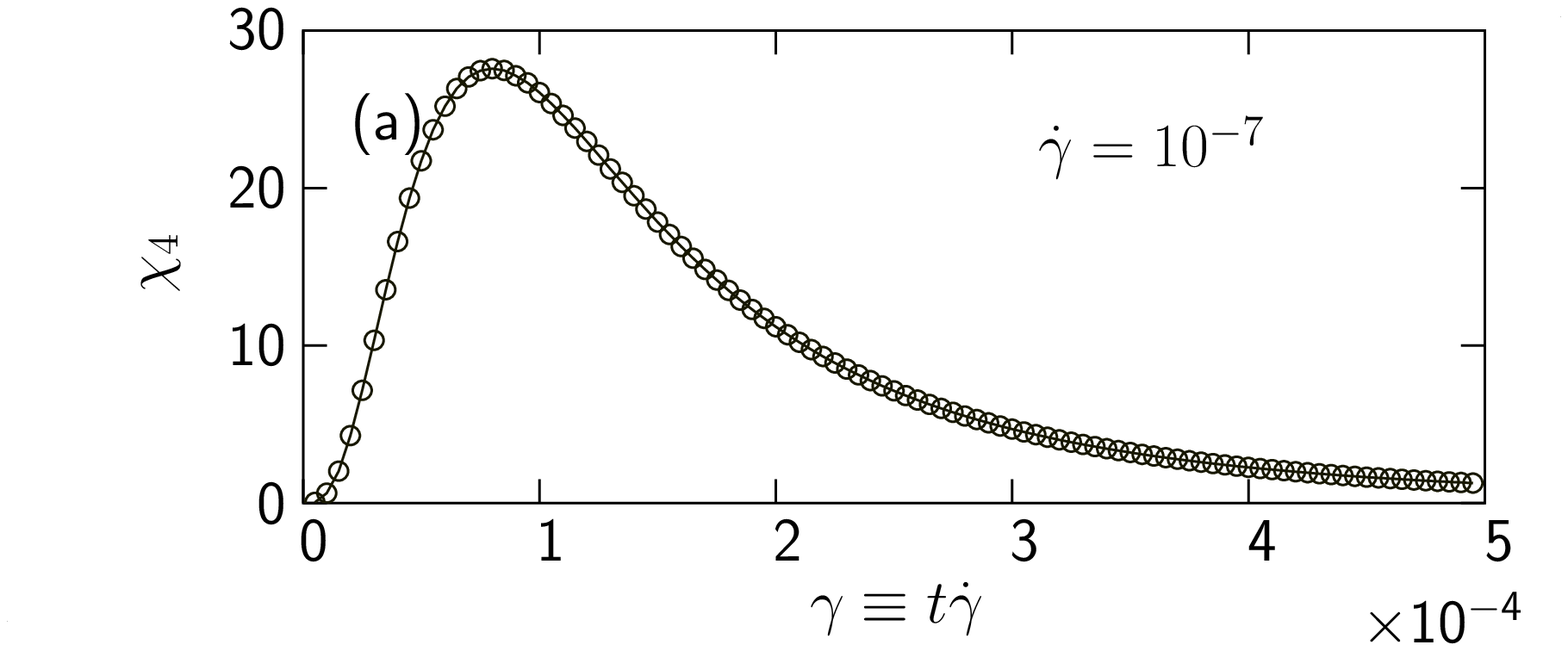}
  \includegraphics[bb=41 319 532 545, width=7cm]{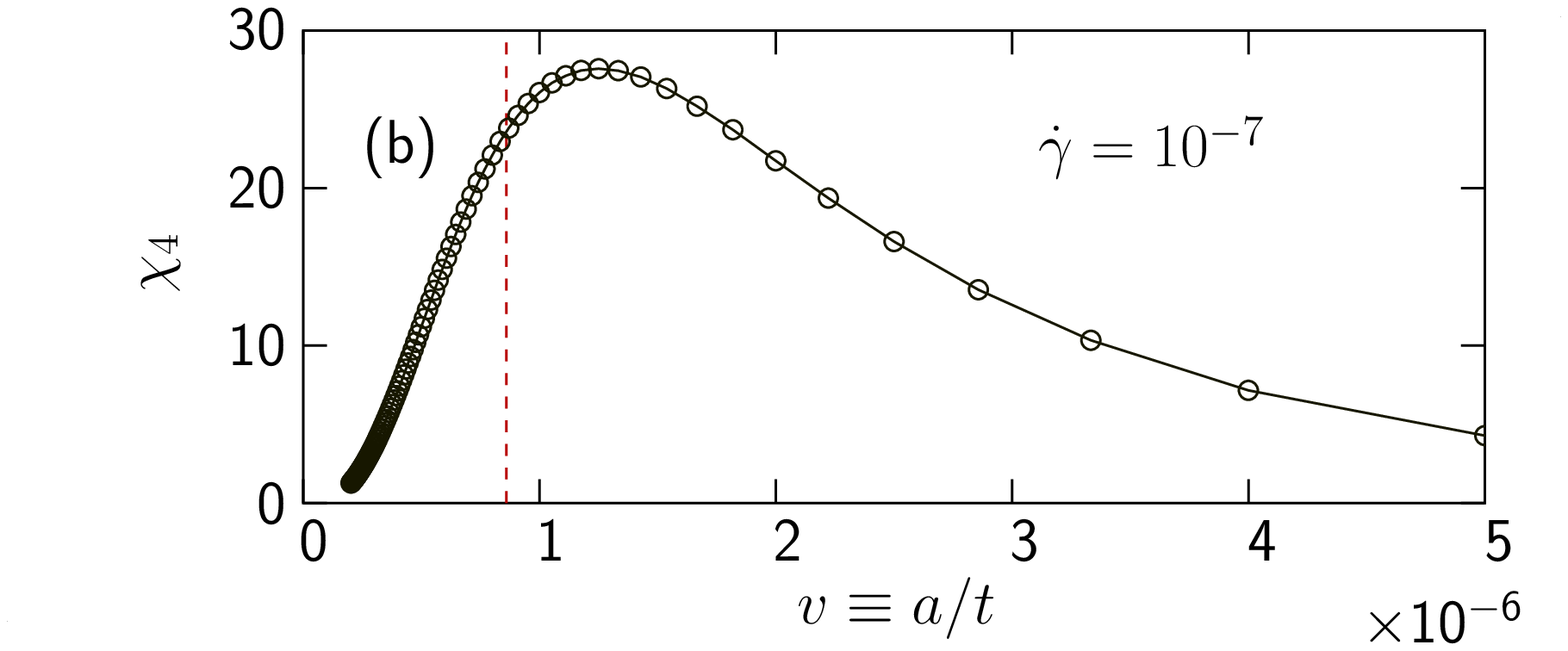}
  \caption{Dynamical susceptibility determined with probing length $a=0.001$. Panel (a) is
    $\chi_4$ vs $\gamma$ determined from the fluctuations in $Q_1(a,\gamma)$ which is, in
    turn, essentially the fraction of particles that have moved the distance $a$ during
    the shear $\gamma$. Panel (b) is the same data but plotted against
    $v=a/t\equiv a\gdot/\gamma$, which is the average velocity needed for the particle to
    move the distance $a$ during a shear $\gamma$. We note that peak in $\chi_4$ is not
    far from the peak velocity $v_p$, shown by the dashed line.}
  \label{fig:chi4}
\end{figure}

To demonstrate that the velocity correlations are dominated by the slow particles we first
examine the overlap function \cite{Lechenault_2008, Heussinger_Berthier_Barrat:2010}
which for each individual configuration is determined from the positions of particles $i$
at a reference time $\rr_i(0)$ and the positions at a time $t$ later, but compensated for
the affine displacement, i.e.\ $\rr_i(t)-\Delta_i(t)\hat x$. The overlap function is then
\begin{displaymath}
  Q_1(a, t) = \frac 1 N \sum_{i=1}^N \exp\left(-\frac{|\rr_i(t)-\Delta_i(t)\hat x-\rr_i(0)|^2}{2a^2} \right),
\end{displaymath}
where $a$ is a probing distance.  The affine displacement, from the affine velocity field,
$v_x=y\gdot$, is given by $\Delta_i(t) = \int_0^t y_i(t')\gdot dt'$. The dynamic
susceptibility is \cite{Heussinger_Berthier_Barrat:2010}
\begin{equation}
  \chi_4(a, t) = N(\expt{Q_1^2(a, t)} - \expt{Q_1(a, t)}^2).
  \label{eq:chi4}
\end{equation}

\Fig{chi4}(a) shows $\chi_4$ vs $\gamma\equiv t\gdot$. The peak in the plot shows the
amount of shear at which half the particles have moved at least the probing length,
$a=0.001$.  We note that it is possible to extract a typical velocity from this, and
determine the velocity from $v\equiv a/t$. These data are shown in \Fig{chi4}(b) and lead
to the conclusion that the collective dynamics is dominated by particles with
$v_4\approx1.25\times10^{-6}$. We note that this velocity is not far from the peak
velocity, $v_p=0.86\times10^{-6}$, that characterizes the distribution of slow particles.

To show that most of the dissipation---and thus the dominant contribution to the shear
stress--- is due to particles with $v>v_4$, i.e.\ particles with considerably higher
velocities than this characteristic velocity, we note that $S(v_4)/\sigma$---the fraction
of the dissipation due to particles with $v\leq v_4$---is small and decreases with
decreasing shear strain rate. For $\gdot=10^{-6}$, $10^{-7}$, and $10^{-8}$ the respective
fractions are $S(v_4)/\sigma\approx 0.051$, $0.044$, and $0.024$. The conclusion is thus
that correlations and the contribution to the shear viscosity (i.e.\ dissipation) decouple
in the $\gdot\to0$ limit as they are governed by different sets of particles.

A different way to reach the same conclusion is through analyses of the correlation
function \cite{Olsson_Teitel:jam-xi-ell}
\begin{equation}
  \label{eq:g}
  g(x) = \frac{\expt{v_x(0) v_x(x\hat x)} - \expt{v_y(0) v_y(x\hat x)}}{\mathbf{v}^2/2}.
\end{equation}
In \REF{Olsson_Teitel:jam-xi-ell} it was concluded that $g(x)$ may be fitted to
\begin{equation}
  \label{eq:gxAB}
  g(x) = Ae^{-x/\xi} - Be^{-x/\ell},\quad A, B>0,
\end{equation}
where the two terms describe the fluctuations in the rotation and the divergence of the
velocity field. It was furthermore found that the diverging $\eta_p\equiv p/\gdot$ scales
with $\xi$, which thus suggests that it is $\xi$, which describes the decay of the
rotations in the velocity field, that is the more significant correlation length, even
though $\ell$ is often considerably bigger \cite{Olsson_Teitel:jam-xi-ell}.

\begin{figure}
  \includegraphics[width=8cm]{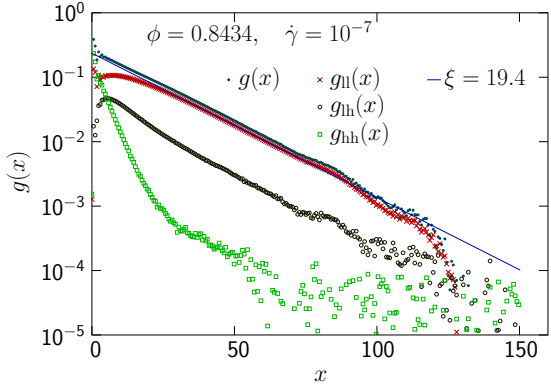}
  \caption{The splitting of the correlation function into three different terms. We here
    designate each particle as having ``low'' or ``high'' velocity with the threshold
    $v_{50}=5.46\times10^{-6}$ chosen such that the sets of particles with low and high
    velocities each dissipate half the power. This is thus similar in spirit to the
    separation into slow and fast particles.  Since each term that contributes to $g(x)$
    involves two particles the full correlation function $g(x)$ may be split into three
    functions: $g_\mathrm{ll}(x)$ from two low velocity particles, $g_\mathrm{lh}(x)$ from
    one low velocity particle and one high velocity particle, and $g_\mathrm{hh}(x)$ from
    two high velocity particles. Since it is $g_\mathrm{ll}$ and (to a less extent)
    $g_\mathrm{lh}$ that dominate the correlations, the conclusion is that it is the low
    velocity particles that are behind the long range correlations in $g(x)$. The solid
    line is $\sim e^{-x/\xi}$ with $\xi=19.4$. The figure is for $N=65536$ particles,
    $\phi=0.8434$, and $\gdot=10^{-7}$.}
  \label{fig:g_lh}
\end{figure}

Since $g(x)$ gives clear evidence for long range velocity correlations it can be used to
demonstrate that the correlations are dominated by the slower particles.  To this end we
define a threshold velocity $v_{50}$ such that half the power is dissipated by particles
with low velocities, $v<v_{50}$ and half by the high velocity particles, $v>v_{50}$. We
thus take $v_{50}$ to be the limit between low and high velocities, which is similar in
spirit to ``slow'' and ``fast'' particles above, but with the difference that there is no
sharp limit in the latter definition as the slow and the fast distributions overlap each
other over a sizable velocity region.  We then split $g(x)$ into terms $g_\mathrm{ll}(x)$,
$g_\mathrm{lh}(x)$, and $g_\mathrm{hh}(x)$, which are the contributions to the correlation
function from two low velocity particles, one particle with low velocity and one with
high, and two high velocity particles, such that the full correlation function is
$g(x)=g_\mathrm{ll}(x) + g_\mathrm{lh}(x) + g_\mathrm{hh}(x)$. These different terms,
obtained at $\phi=0.8434\approx\phi_J$ and $\gdot=10^{-7}$ with
$v_{50}=5.46\times10^{-6}$, are shown in \Fig{g_lh}.

The conclusion from this figure is that it is the low velocity particles that strongly
dominate the correlations. The contributions from $g_\mathrm{ll}(x)$ is about 85\%, from
$g_\mathrm{lh}(x)$ the contribution is about 14\%, and the contribution from
$g_\mathrm{hh}(x)$---two high velocity particles---is less than 1\% at large distances.
In a sense this finding is not surprising since one can expect the build up of long range
correlations in a system of elastic particles to be a slow process whereas the high
velocities only exist for shorter times.

\begin{figure}
  \includegraphics[width=7cm]{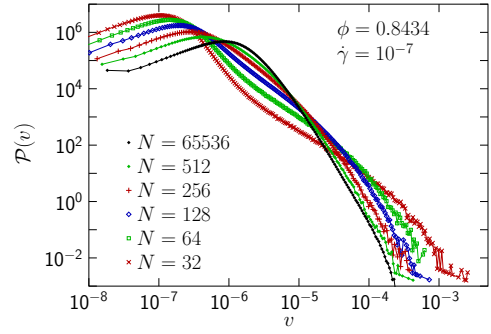}
  \caption{Finite size dependence of $\cP(v)$ at $\phi=0.8434\approx\phi_J$ and
    $\gdot=10^{-7}$. This figure shows that the low velocity region and the high velocity
    region are affected differently by a reduced system size such that the low-velocity peak
    moves to even lower velocities whereas the high-velocity tail extends to higher
    velocities.}
  \label{fig:fss-8434}
\end{figure}

The finding that slower particles contribute more to the velocity correlations than the
faster particles leads to the expectation that a reduced system size should affect
different parts of the velocity distribution differently. This expectation is borne out in
\Fig{fss-8434} where it is found that the peak in the distribution moves to lower
velocities as $N$ decreases whereas the tail moves in the opposite direction to higher
velocities. An attempted explanation of the finite size dependence on the peak velocity is
given in \Sec{ratio}, but we here present an explanation of the shift of the tail in the
distribution to higher velocities. The reasonable explanation is that a reduced system
size means a hindering of certain large-scale reorganizations that are needed for finding
new low-energy configurations. When these large-scale reorganizations are no longer
possible the system builds up bigger tensions, which are now and then reduced in more
dramatic events with higher velocities, which leads to a shift of the tail of the velocity
distribution to higher velocities.

\subsection{Attempts to rationalize the findings}
\label{sec:ratio}

As an attempt to rationalize the findings we start by considering the slow process and
turn to the fast process as a second step.

As a starting point we consider two contacting hard particles initially at rest at
different $y$ coordinates, $\pm y/2$ and separation $d_0\mathbf{n}$ with the unit vector
$\mathbf{n}=(n_x, n_y)$. Due to the homogeneous velocity profile these particles will
experience opposite forces from this flow along the $x$ direction,
$\pm (y/2) k_d \gdot\hat x$, and also contact forces $\fel_\pm$ in direction
$\pm\mathbf{n}$. If there are no other interacting particles, the total velocities
$\vtot_\pm$ will be $\vtot_\pm \mp y\gdot\hat x = \fel_\pm/k_d$, which together with
$\f^\mathrm{el}\parallel\mathbf{n}$ and $\v^\mathrm{tot}\perp\mathbf{n}$ gives
\begin{eqnarray*}
  n_y\vtot_\pm & = & n_x\fel/k_d \pm (y/2)\gdot,\\
  n_x\vtot_\pm & = & -n_y\fel/k_d,
\end{eqnarray*}
and the relative particle velocity 
\begin{displaymath}
  \vtot \equiv \vtot_+-\vtot_- = n_y y\; \gdot.
\end{displaymath}
In the presence of other particles that could hinder the displacement we expect this to
instead lead to a force $k_d v^\mathrm{tot}$. Since the velocities at higher densities are
correlated across a distance $\xi$ \cite{Olsson_Teitel:jam-xi-ell} it follows that any
given contact should contribute a quantity $\propto\gdot$ to the velocity field of each
particle in the volume $\sim\xi^2$ centered at that contact.

We now instead turn to the behavior of a single particle and a consequence of the above
discussion is that its velocity becomes affected by the $n=c_\xi^2 \xi^2/d_0^2$ contacts
in a volume $\xi^2$, where $c_\xi$ is a factor of order unity. We further assume that the
relative velocity $v^\mathrm{tot}_k\sim d_0\gdot$ at contact $k$ contributes
$\bm\eta_{ik}d_0\gdot$ to the velocity of particle $i$.  For simplicity we take
$\bm\eta_{ik}$ to be random and independent with $\expt{\bm\eta_{ik}}=0$ and
$\expt{\bm\eta_{ik}^2}=c^2_\eta$. The velocity of a given particle then becomes
$\v_i = \sum_{k=1}^n \bm\eta_{ik} d_0\gdot$ where the sum is over the $n$ contacts with
$\rr_{ik}<\xi$. This gives $\expt{\v_i}=0$, and the variance
$\expt{\v_i^2}=n c^2_\eta d_0^2 \gdot^2$ then defines a characteristic velocity
\begin{equation}
  \label{eq:vprim_xi}
  v' = \sqrt{\expt{\v_i^2}} = c_\eta\sqrt n d_0 \gdot = c\gdot\xi,
\end{equation}
where $c\equiv c_\eta c_\xi$ is a constant of order unity.  For hard disks (or equivalently,
soft disks at $\gdot\to0$) at densities below $\phi_J$ this becomes (cf.\ \Eq{power})
\begin{equation}
  \eta'_\mathrm{hd} = \frac N V k_d \frac{v'^2}{\gdot^2} = \frac N V k_d c^2 \xi^2 \sim \xi^2,
  \label{eq:eta_xi}
\end{equation}
and together with $\xi\sim(\phi_J-\phi)^{-1}$ \cite{Olsson_Teitel:jam-xi-ell} this leads to
\begin{equation}
  \eta'_\mathrm{hd} \sim (\phi_J-\phi)^{-2},
  \label{eq:eta_slow}
\end{equation}
which is an estimate of the contribution from the slow particles, only, and not the
full shear viscosity.

For an order of magnitude check we turn to low densities $\phi=0.78$ through 0.83 where
the contribution from the slow particles should dominate the total $\eta$, determine $\xi$
as in \REF{Olsson_Teitel:jam-xi-ell} and make use of values of $\eta$ together with
\Eq{eta_xi} to determine
\begin{displaymath}
  c^2 = \frac{\eta}{k_d (N/V) \xi^2} = 0.8\pm0.2.
\end{displaymath}
which shows that $c$ is indeed a constant of order unity.

After this discussion of hard particles below jamming we turn to the behavior at $\phi_J$.
We then make use of the correlation length $\xi\sim\gdot^{-1/z}$, with $1/z=0.26$
\cite{Olsson_Teitel:jam-xi-ell}.  \Eq{vprim_xi} then gives the characteristic velocity
\begin{displaymath}
  v' \sim \gdot\;\gdot^{-1/z} \sim \gdot^{u'},
\end{displaymath}
with the exponent
\begin{displaymath}
  u' = 1-1/z = 0.74,
\end{displaymath}
which is very close to $u_v=0.766$ for the peak velocity, $v_p\sim\gdot^{u_v}$ in
\Eq{uv}. Though this agreement is encouraging as it suggests a connection between very
different quantities, we note that the reasoning is still very incomplete as the behavior
of $\xi$ is taken as a given starting point without any motivation.

\begin{figure}
  \includegraphics[width=7cm]{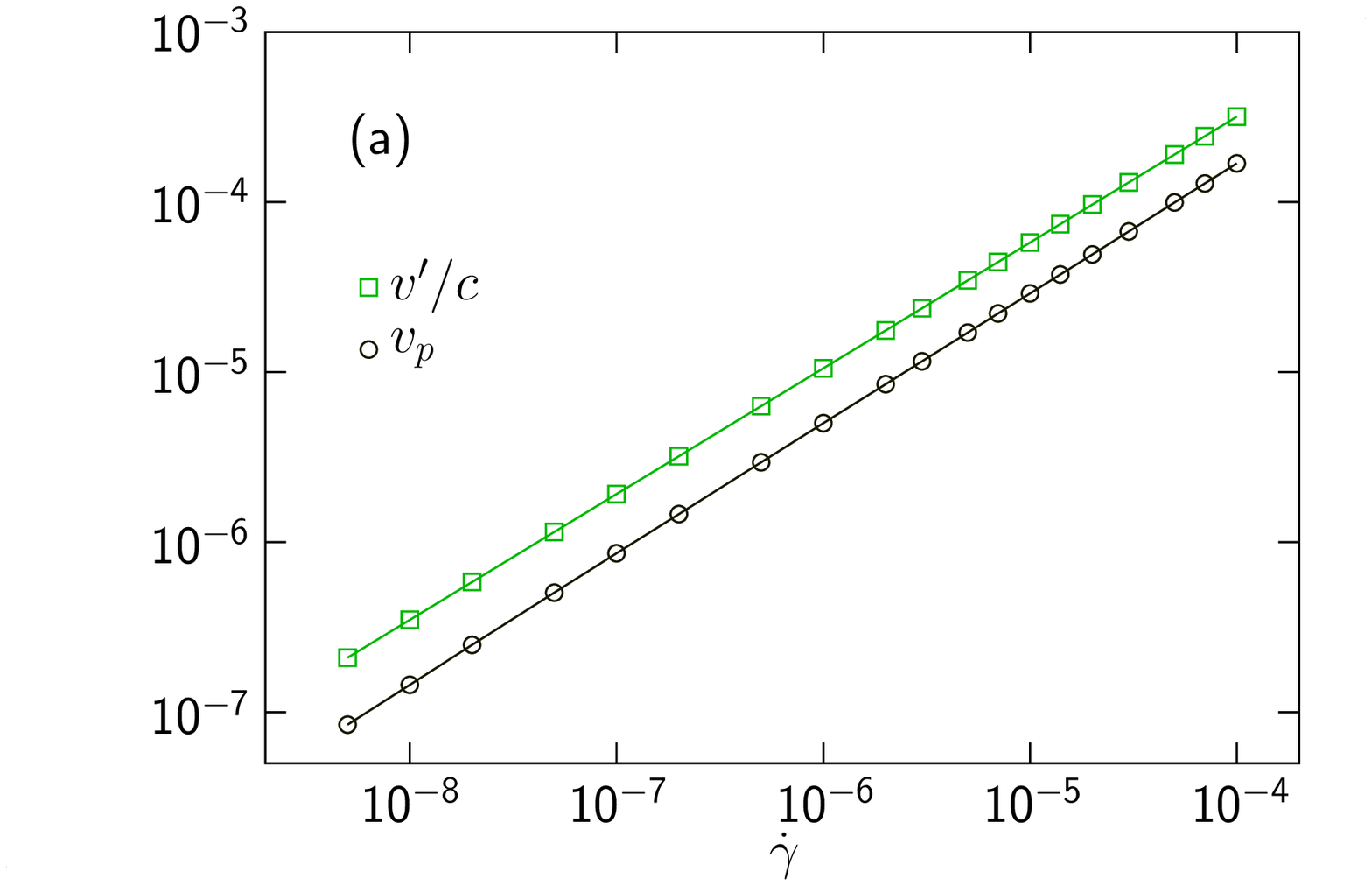}
  \includegraphics[width=7cm]{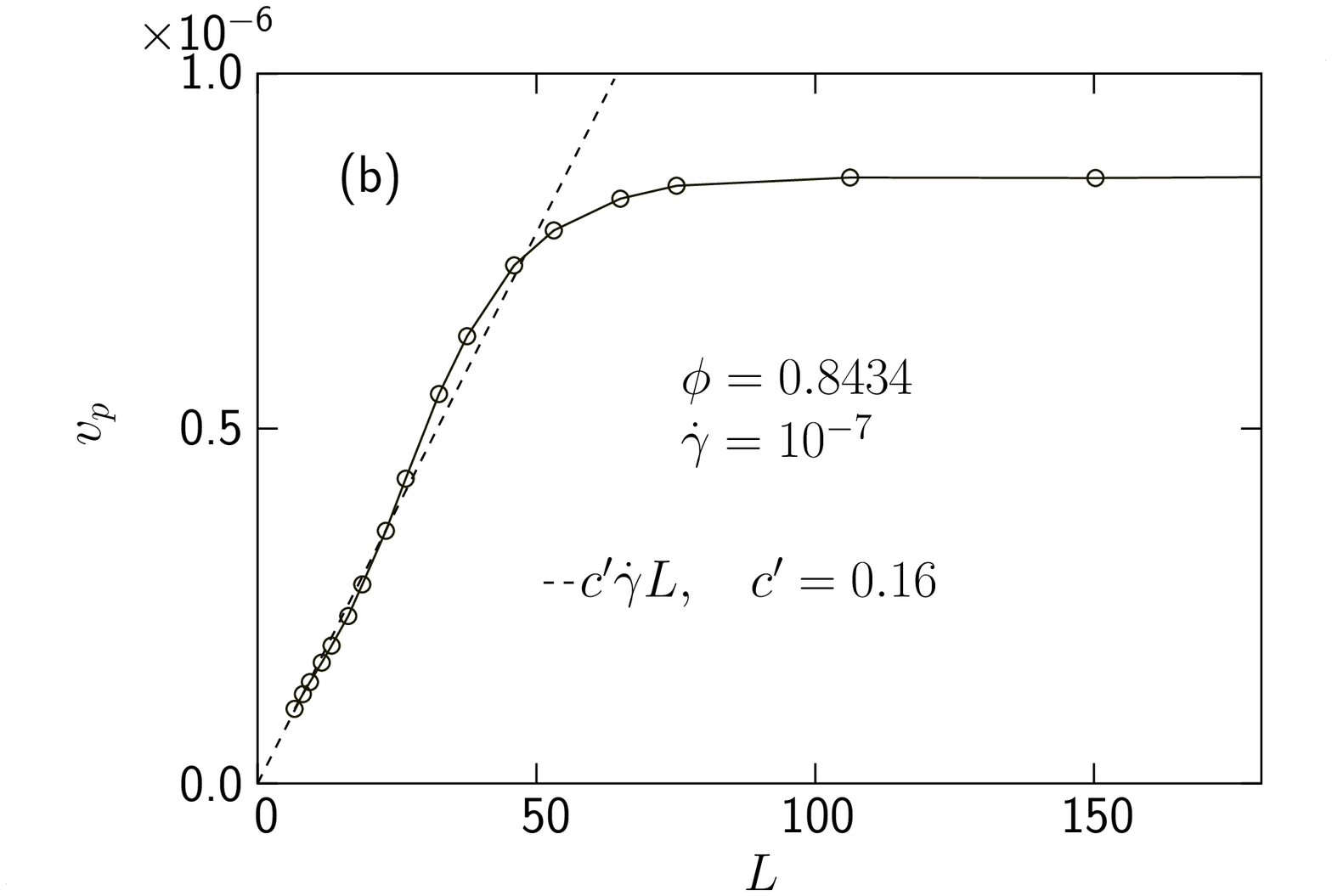}
  \caption{Attempts to test the rationalization of the shear rate dependence of the peak
    velocity in \Eq{vprim_xi}. Panel (a) shows a comparison between the peak velocity,
    $v_p$, and the characteristic velocity, from \Eq{vprim_xi}, shown as $v'/c$; the data
    are encouragingly similar. (The open squares are the values of $v'$, assuming
    $c=1$. Taking $v'=v_p$ at $\gdot=10^{-7}$ gives $c=0.45$.) Panel (b) shows the finite
    size effect on the peak velocity, $v_p$, by plotting $v_p$ vs $L$. The linear behavior
    $c'\gdot L$ with $c'=0.16$, at small $L$, shown by the dashed line, is consistent with
    predictions in the main text. (The correlation length at $\phi\approx\phi_J$ and
    $\gdot=10^{-7}$ is $\xi\approx 19$.)}
  \label{fig:vp-L-e100}
\end{figure}

\Fig{vp-L-e100}(a) shows a direct comparison of $v_p$ and $v'/c$ using
$\xi\approx 0.29\gdot^{-1/z}$ \cite{Olsson_Teitel:jam-xi-ell} in \Eq{vprim_xi}, and we
note that they are very similar.  The points $v'/c$ are simply the values of $v'$ when
taking the unknown constant to be $c=1$.

[As a digression we now return to the behavior of hard particles below $\phi_J$ to compare
our predictions based on $\sigma_s$ with \Eq{eta_slow}. From the very similar behaviors of
$v'$ and $v_p$ one could expect an excellent agreement between predictions from $\sigma_s$
and \Eq{eta_slow}, but there is instead a clear difference. For this discussion we make
use of $\beta_2$, introduced in \Sec{hard-particle-limit}, for the divergence of the
secondary term. With $q_2=0.567$ and $z\nu=1/0.26$ $\beta_2=(1-q_2)z\nu\approx1.67$, is
quite different from $\beta_2=2$ in \Eq{eta_slow}. Recalling \Eq{uw} and $q_2=u_w$ it
turns out that one way to get $\beta_2=2$ is if the equalities $u_v=1-1/z\nu$ (this is
$u_v=u'$) and $u_v+u_\cP=0$ were both fulfilled, but since they are only approximately
fulfilled, the exponent instead becomes somewhat lower.  It is interesting to note that
$u_v+u_\cP=0.033>0$ means that the fraction of particles with velocities up to the peak
increases slowly with decreasing $\gdot$. Such a trend is possible only because of the
existence of two different processes.]

It is most interesting to also examine the dependence on system size. The starting point
is then that a quantity which is determined from processes in a correlation volume should
have a finite size dependence unless the linear system size is $L\gg\xi$. For small $L$
one expects $L$ to take the place of $\xi$, and \Eq{vprim_xi} then becomes
$v' \sim\gdot L$. \Fig{vp-L-e100}, which shows $v_p$ vs $L$, gives evidence for such a
behavior as the data below $L\approx50$ follow the dashed line, $c'\gdot L$, to a good
approximation. This is also the likely explanation of the size-dependence of $v_p$ in
\Fig{fss-8434} which is $v_p\sim L$ for $N\leq512$.

Even though this picture describes the slow process, only, it also holds the seed to the
fast process that gives particles with considerably higher velocities. We first recall
that the condition for a wide tail in the velocity distribution is the presence of large
contact forces, i.e.\ that the typical contact force is considerably larger than the
typical net force $k_d v'$ that drives the slow particles. The typical contact force,
$f'$, may be determined from the pressure which is given by $p'=\sigma'/\mu$, (where $\mu$
is the dimensionless friction). From $V/N\approx d_0^2$, $p'\approx \frac 1 4 f' z/d_0$
and the approximate expressions for the contribution to the shear stress from the slow
particles,
\begin{displaymath}
  \sigma' \approx \frac N V \frac{k_d}{\gdot} v'^2,
\end{displaymath}
and \Eq{vprim_xi} one finds
\begin{equation}
  f' = \frac N V \frac 1 \mu \frac{k_d}{\gdot}d_0 v'^2 \approx \frac{c}{\mu} \frac{\xi}{d_0}k_d v',
\end{equation}
for the typical contact force. In most cases the contact forces on a particle almost
cancel each other out, but in the case where the forces fail badly to balance each other
out one finds
\begin{equation}
  \label{eq:vfast}
  v_\mathrm{fast} = c_g f'/k_d = \frac{c_g c}{\mu} \frac{\xi}{d_0} v',
\end{equation}
and even though the geometrical factor is $c_g\ll 1$, a big $\xi$ together with
$1/\mu\approx 10$ (which holds close to jamming) may lead to velocities
$v_\mathrm{fast}\gg v'$. (That $c_g\ll 1$, is illustrated in \Fig{snapshot}(a) where the
three particles are almost in a line and therefore give a resultant force that is
considerably smaller than the contact forces.)

What finally gives the very high velocities, with tails extending up to $v\approx100\;v_p$
for $\gdot=10^{-7}$, is the fact that the above mentioned mechanism is self-amplifying
since a number of fast particles have the effect to make $\expt{v^2}>v'^2$, which then
increases $\sigma$ and the typical force, which in turn has the effect to increase
$\expt{v^2}$ even more.

\section{Discussion}

\emph{Short summary:} The study of the velocity distribution in the present paper suggests
the existence of two different processes with different scaling properties. We call them
the slow process and the fast process as they are dominated by the slower particles in the
peak and the faster particles in the tail of the distribution, respectively. Due to the
relation between input power $\sigma\gdot$ and dissipated power $k_d\expt{v^2}$,
\Eq{power}, the shear stress is thought of as being controlled by the dissipation, which
makes it possible to split the shear stress into contributions from the slow process and
the fast process, $\sigma=\sigma_s+\sigma_f$. It is then found that the leading divergence
of the shear viscosity is governed by the fast process whereas the correction-to-scaling
term from the critical scaling analysis is related to the slow process. Since it is
furthermore found that the long range velocity correlations that develop as criticality is
approached, are due to the slow process, it appears that the connection expected in
critical phenomena between the diverging correlation length and the diverging viscosity,
is an indirect one, only. Taken together this suggests that shear-driven jamming is an
unusual kind of critical phenomenon.

\emph{Open questions:} There remain several open questions and one of them is on the
mechanism behind the algebraic velocity distribution in the fast process. Since
$v_i=f^\mathrm{el}_i/k_d$ the velocities, and thereby the velocity distribution, are
directly given by the sum over the contact forces,
$\f^\mathrm{el}_i=\sum_j \f^\mathrm{el}_{ij}$. The contact forces $\f^\mathrm{el}_{ij}$
are here from a narrow distribution whereas the distribution of the velocities (through
the net forces) have a tail, $\sim v^{-r}$, with different $r$. An open question is what
mechanism there is that generates this distribution.

A related enigmatic finding is that the values of $q$ and $q_2$ together give
$q_2/q=1.995\pm0.021$ (three standard deviations) which suggests the simple relation
$q_2/q=2$. Though $q_2$ may be ``understood'' from the dependence of the velocity
distribution on $\gdot$, there is no simple way to come to grips with the exponent $q$
since it depends on both the exponent $r$, which changes with $\phi$ and $\gdot$, and
other properties of the tail of the distribution, in an opaque way. We here just speculate
that there is a coupling between the two different processes that makes the system adjust
itself to give this simple relation between the slow and the fast processes, but we have
no clue to the underlying mechanism.

In critical phenomena the behavior is largely controlled by the main term, but in view of
the present findings, that the diverging correlations appear to be present in the slow
process, only, it could be that it is rather the slow process that is central in the
critical phenomenon and, in some way, controls the fast process. If this is so it is
perhaps more appropriate call $\sigma_2$ in \Eq{fit-sigma} the ``secondary term'' rather
than the correction-to-scaling term, as the latter term has the strong connotation of
being small and insignificant.

\emph{Bucklers and the dimensionality:} The fast particles in \Fig{snapshot}(a) are
similar to the bucklers described in \REF{Charbonneau:2015:prl} which are found to be
related to localized excitations. From that work it is also known that the population of
bucklers decreases with higher dimensions and one could expect that this should also mean
a lower frequency of fast particles and perhaps also that this separation into two
different processes would no longer be relevant. We have however done some preliminary
studies of the velocity histograms in both three and four dimensions and it is then clear
that the picture described here remains essentially the same also in these higher
dimensions. This could perhaps suggest that the processes as in \Fig{snapshot}(b),
that give chains of fast particles, could be more important in higher dimensions.

\emph{Contact changes:} Contact change events have been studied through quasistatic
shearing of soft spheres and one has then found that these contact change events are of
two different kinds where the first is irreversible and dramatic ``rearrangements'' that
lead to discontinuous change of positions and the second is reversible and smooth
``network events'' \cite{Morse:prr}. The first kind has also been termed ``jump changes''
whereas the continuous contact change is termed a ``point change''
\cite{Tuckman:SoftMatter:2020}.  It does indeed seem that the fast and slow processes of
the present work are respectively related to these different kinds of contact changes, and
beside adding credibility to our picture of two different process, this connection also
suggests new avenues for further research.

\emph{Relation to theoretically determined exponent:} A further question is the connection
between our findings and the theoretically determined value of the exponent
$\beta/u_z$. The assumption that the process that governs the divergence of the shear
viscosity is ``spatially extended'' \cite{DeGiuli:2015} or ``extensive''
\cite{Harukuni-logcorr:2020}, is in contrast to our finding that the fast particles are
short range correlated, only.  Our finding could suggest going back to
\REF{Lerner-epl:2012} that presented a different results when using $\theta_\ell=0.18$
from the distribution of weak forces (determined for all contacts and not only the
``extended'' ones \cite{DeGiuli:2015, Charbonneau:2015:prl}) and gave the value
$\beta/u_z=(3+\theta_\ell)/(1+\theta_\ell)=2.69$ in excellent agreement with the
simulations in 2D \cite{Olsson:jam-tau}. In spite of this agreement in 2D (which could
perhaps be just fortuitous) a remaining question is the reason for the different exponent
in three dimensions, and we conclude that more work is needed to sort out this question.

\emph{Future and ongoing work:} There are quite a few interesting directions for the
further research. As already mentioned a finite size scaling study of shear-driven
jamming, by means of the splitting into $\sigma_s$ and $\sigma_f$, is under way. We then
also plan to examine models with elliptical and ellipsoidal particles, and/or with
different models for dissipation, with the key question what properties of the model that
determine the universality class of the transition. It would also be interesting to
examine how the introduction of inertia---which is known to give an altogether different
behavior \cite{Trulsson:2012}---is reflected in the properties of the velocity
distribution.

\begin{acknowledgments}
  I thank S. Teitel for many illuminating discussions. The computations were enabled by
  resources provided by the Swedish National Infrastructure for Computing (SNIC) at High
  Performance Computer Center North, partially funded by the Swedish Research Council
  through grant agreement no.\ 2018-05973.
\end{acknowledgments}

\appendix

\section{Determination of $\phi_J$ and the exponents $q$ and $q_2$}
\label{sec:q_q2}

\begin{figure}
  \includegraphics[width=7cm]{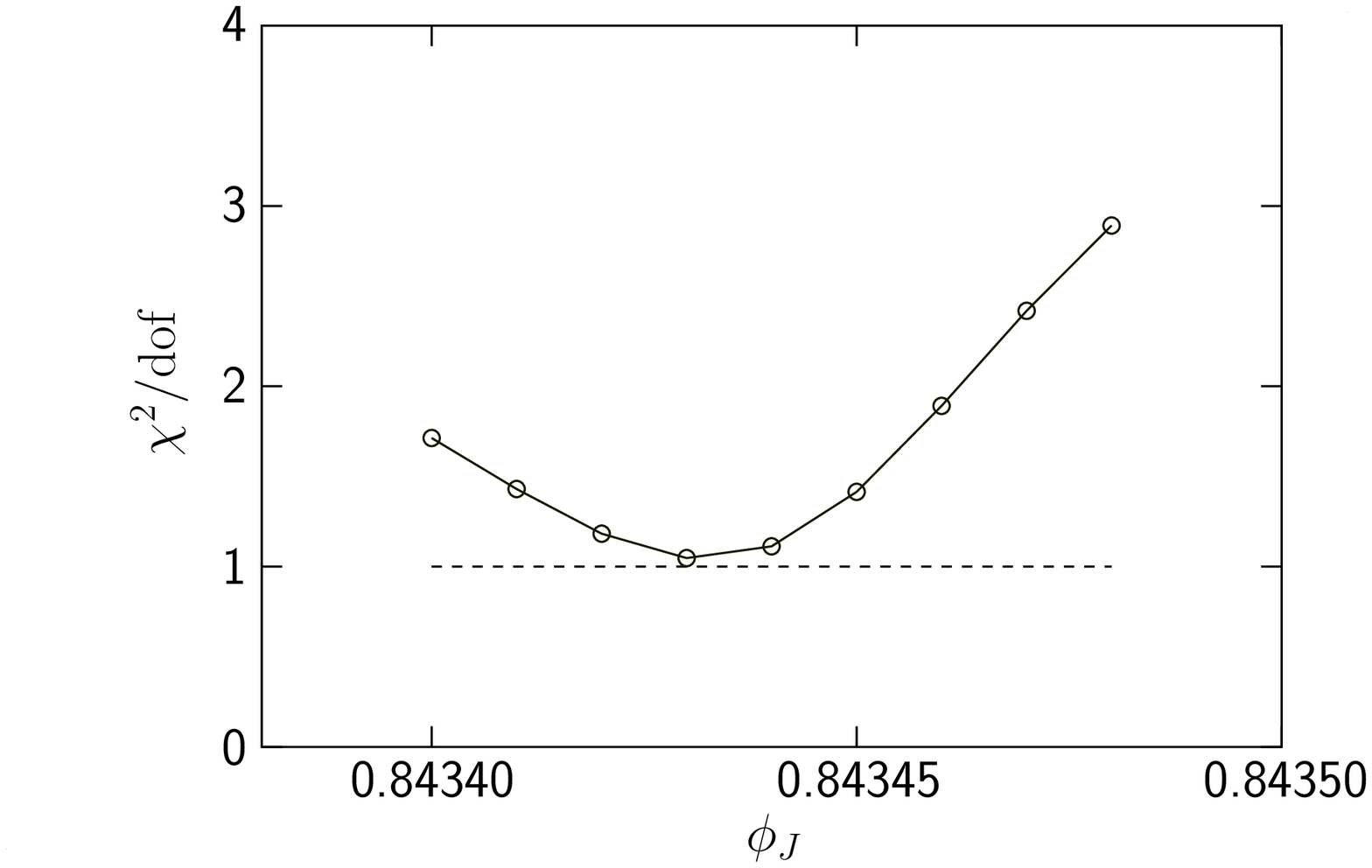}
  \caption{Determination of the jamming density. The figure shows the quality of the fits
    in terms of $\chi^2/\mathrm{dof}$ when assuming different values of $\phi_J$ and using
    method C, i.e.\ demanding that both $q$ and $q_2$ should be the same in the fit of
    $\sigma(\phi,\gdot)$ to \Eq{fit-sigma} and in the fit of $p(\phi,\gdot)$ to
    \Eq{fit-p}. The value $\phi_J\approx 0.843~43$ obtained here was used in the
    determination of $I_2$ shown in \Fig{est-I2}, since that determination is very
    sensitive to the value of $\phi_J$; we have otherwise used $\phi_J\approx0.8434$
    throughout the paper.}
  \label{fig:chisq-phiJ}
\end{figure}

To determine the exponents with the highest possible precision we simultaneously fit shear
stress to \Eq{fit-sigma} and pressure to \Eq{fit-p}.  We are then inspired by
\REF{Rahbari_Vollmer_Park} who use the same exponents $q$ for both quantities and
$q^{(p)}_2=q_2$. That $q$ should be the same for both quantities follows from the
understanding that $\mu\equiv\sigma/p$ approaches a constant at jamming, whereas the same
value of the exponent for the second term for both quantities follows from the
correction-to-scaling exponent being the same for different quantities. Just in order to
examine all possibilities we have however also examined the possibility that the secondary
exponents could be different, and in \Tab{q-q2} we therefore show results from a few
different kinds of fits. Method (A) is from fitting $\sigma$ only, method (B) is from a
simultaneous fit of $\sigma$ and $p$ where we take $q$ to be the same for both $\sigma$
and $p$ but let $q_2$ and $q^{(p)}_2$ be different fitting parameters. Since the
correction term is considerably smaller for $p$ than for $\sigma$, the main effect of
including data for $p$ is to get better precision in $q$ which in turn gives a smaller
error in $q_2$. In method (C) we demand $q^{(p)}_2=q_2$ which gives slightly lower values
of both $q$ and $q_2$. The simultaneous fitting of $\sigma$ and $p$ gives a very sensitive
method and \Fig{chisq-phiJ} shows how the quality of the fit depends on the assumed
$\phi_J$. The optimal fit is obtained with $\phi_J=0.843~43$, just slightly higher than
$\phi_J\approx0.8434$ used throughout this paper. We also note that the values are in good
agreement with \REF{Olsson_Teitel:gdot-scale} that gave $\phi_J=0.843~47$, $q=0.28(2)$,
and that our $q_2-q=0.285(5)$ is in good agreement with $\omega/\nu=0.29(3)$
\cite{Olsson_Teitel:gdot-scale}. Just as in \REF{Rahbari_Vollmer_Park} it is the
combination of two sets of data that narrows down the possible values of $\phi_J$ to a
very small interval in $\phi$.

\begin{table}
  \centering
  \begin{tabular}{|c|c|l|l|l|}
    \hline
    method & $\phi_J$ & $q$ & $q_2$ & remark \\
    \hline
    A & 0.8434 & 0.29(2) & 0.58(5) & fitting $\sigma$, only, \\
    B & 0.8434 & 0.290(2) & 0.58(1) & $q^{(p)}_2=1.1(5)$ \\
    C & 0.8434 & 0.284(2) & 0.567(7) & demanding $q^{(p)}_2= q_2$ \\
    \hline
    C & 0.84343 &  0.281(3) & 0.567(8) & at $\phi_J$ from \Fig{chisq-phiJ}. \\
    \hline
  \end{tabular}
  \caption{Four different determinations of the exponents $q$ and $q_2$. Method A which is
    from using $\sigma(\phi=0.8434,\gdot)$ only gives rather poor precision in the
    exponents. In method B we make use of $p(\phi=0.8434,\gdot)$ to give higher precision
    in $q$, but keeping $q^{(p)}_2$ as a separate fitting parameter from $q_2$. In method C
    we demand $q^{(p)}_2= q_2$, but still assume $\phi_J=0.8434$. The last line is from a
    fit with method C but assuming different jamming densities $\phi_J=0.843~40$ through
    0.843~48. From the quality of the fit, shown in \Fig{chisq-phiJ}, we then determine
    $\phi_J=0.843~43$ which is our value of $\phi_J$. In this determination the
    $\sigma(\phi,\gdot)$ are obtained by interpolating $\sigma(\phi,\gdot)$ measured at $\phi=0.8434$
    and 0.8435.}
  \label{tab:q-q2}
\end{table}

\begin{figure*}
  \includegraphics[bb=11 319 532 538, width=7cm]{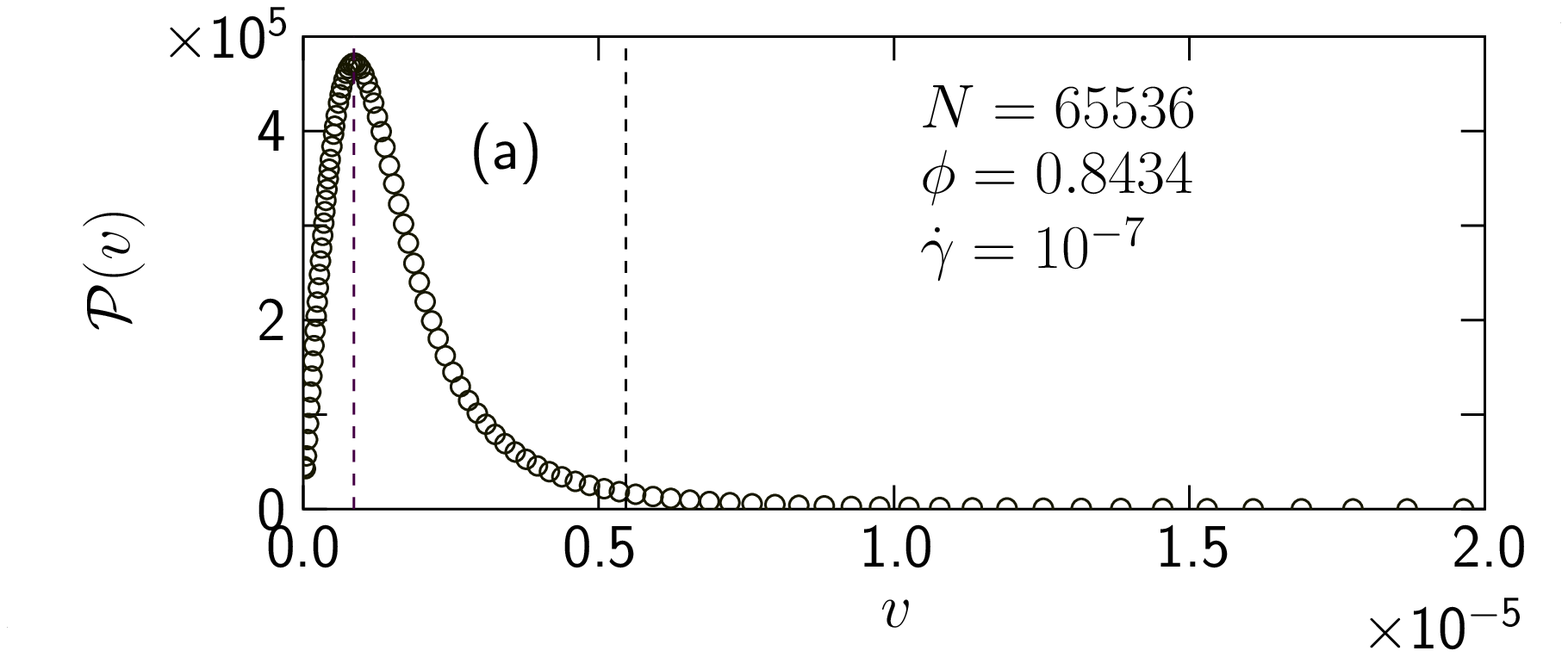}
  \includegraphics[bb=11 319 532 538, width=7cm]{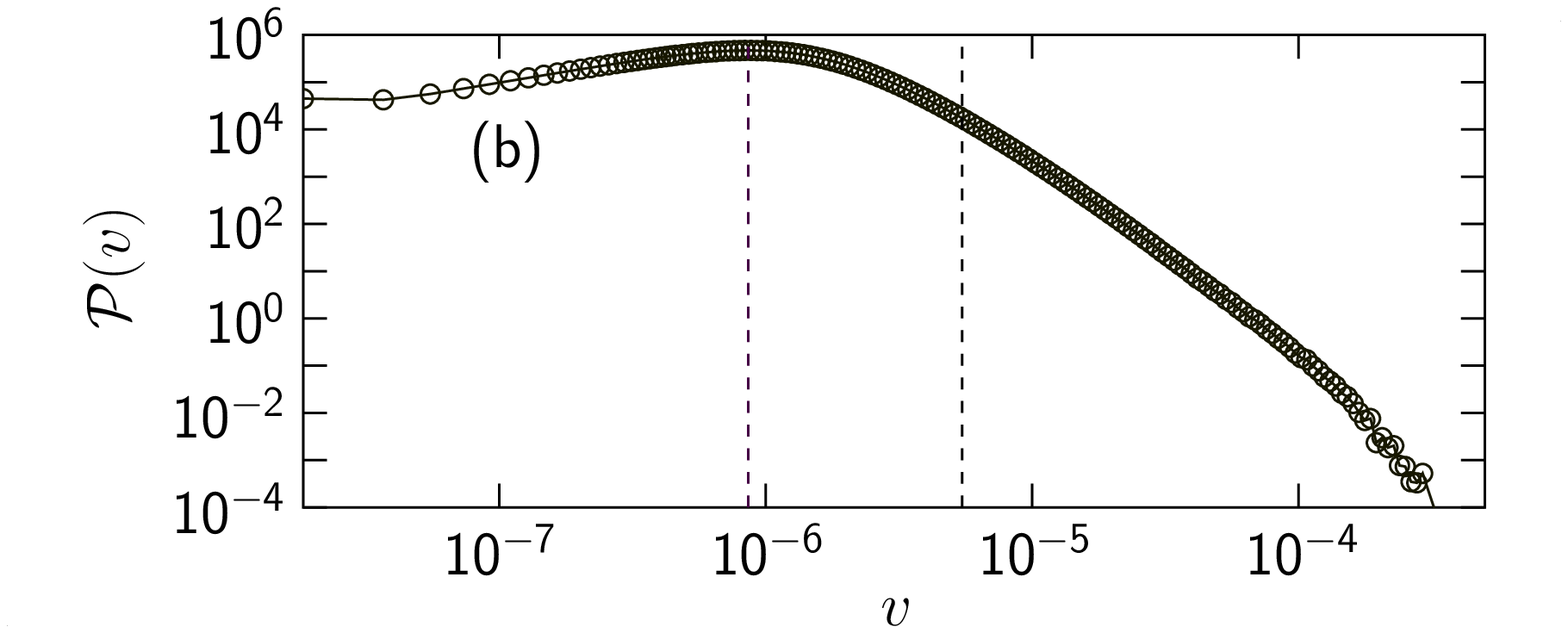}
  \includegraphics[bb=11 319 532 538, width=7cm]{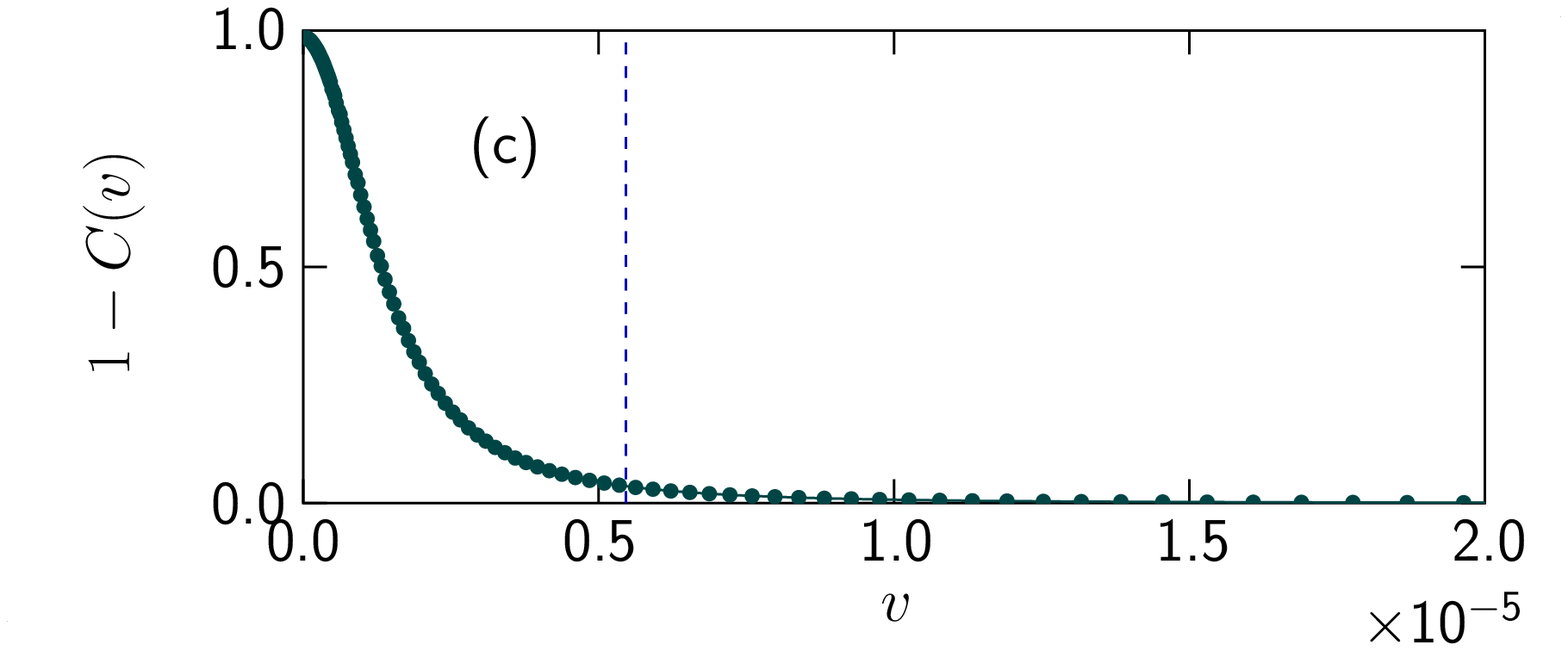}
  \includegraphics[bb=11 319 532 538, width=7cm]{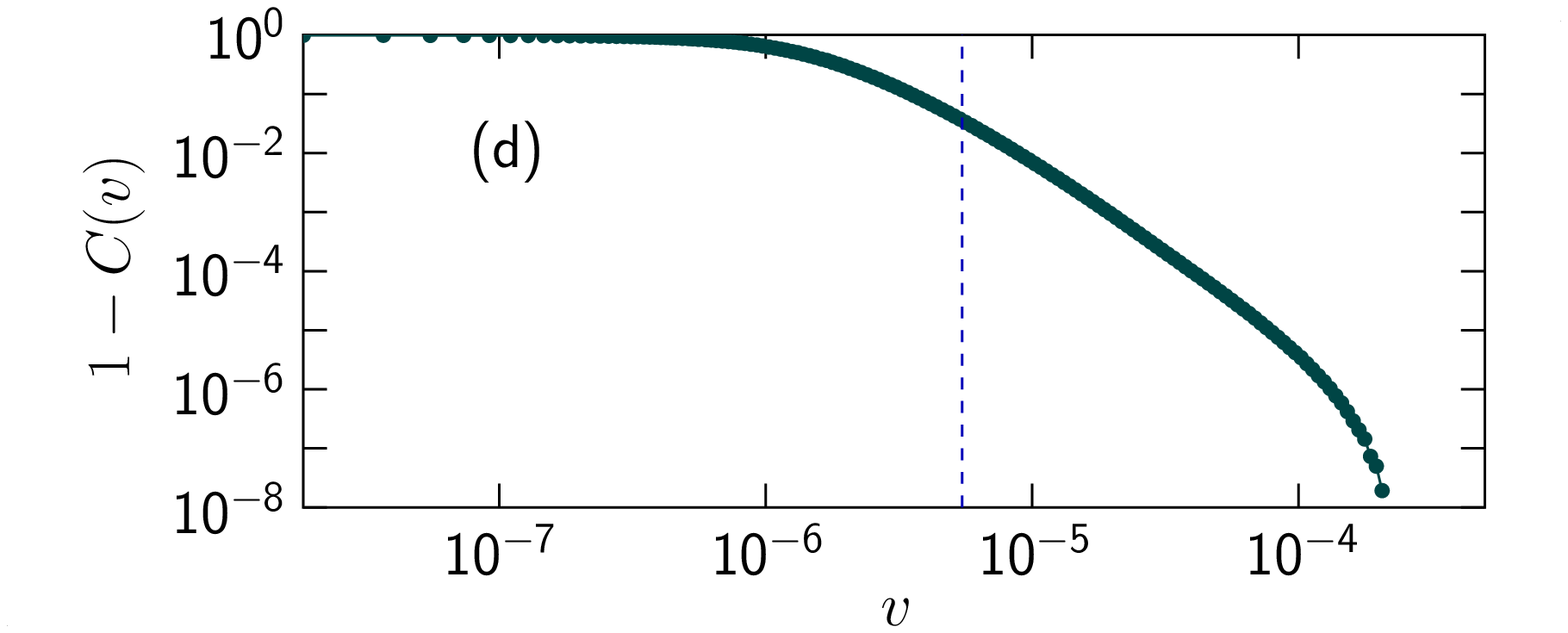}
  \includegraphics[bb=11 319 532 538, width=7cm]{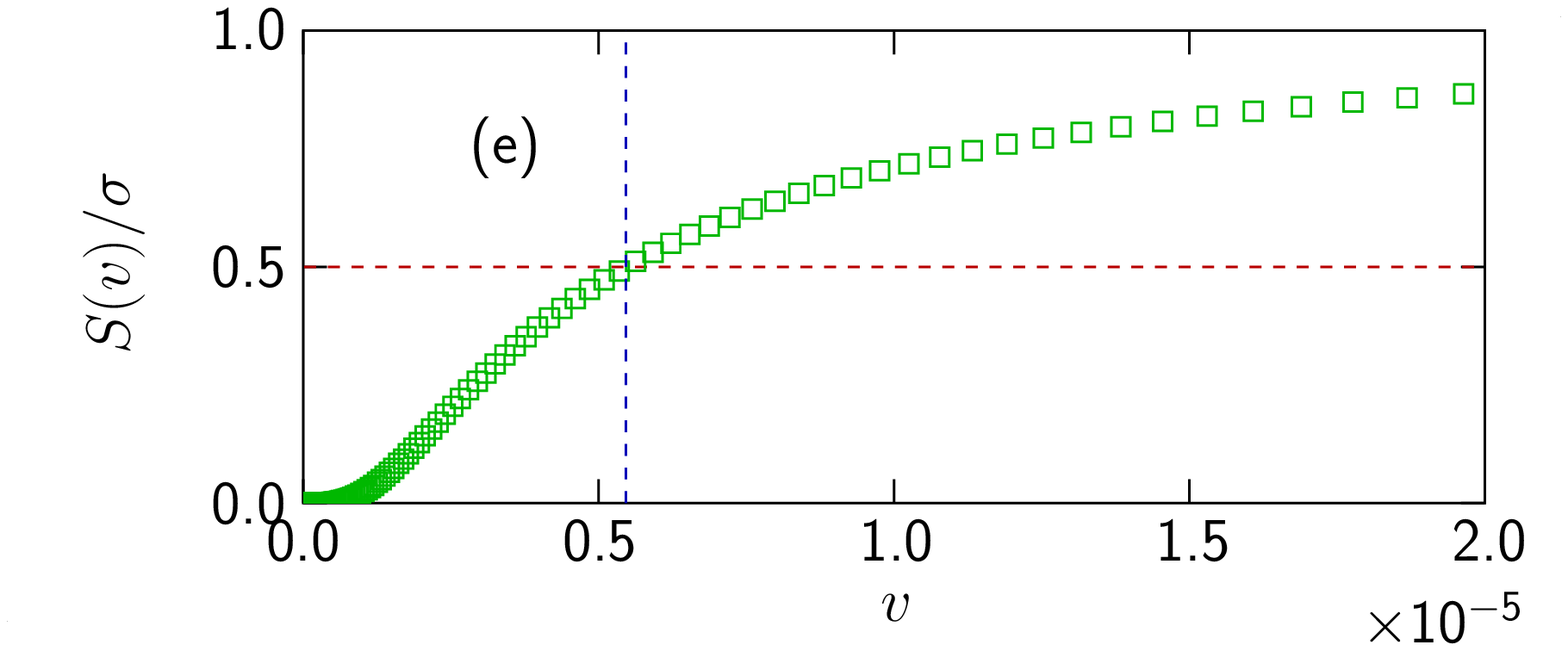}
  \includegraphics[bb=11 319 532 538, width=7cm]{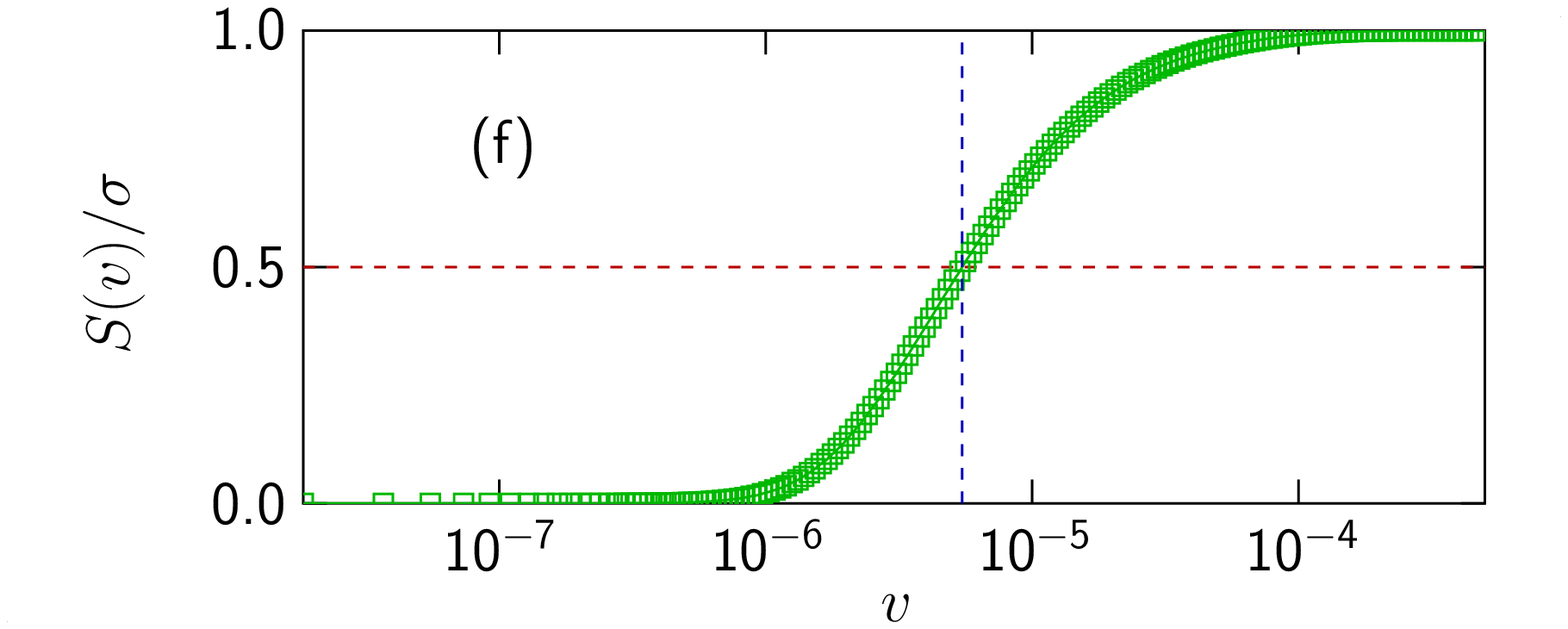}

  \caption{Velocity distribution and relative dissipation at $\phi\approx\phi_J$ and
    $\gdot=10^{-7}$. Panel (a) shows $\cP(v)$ on linear scales (which simplifies the
    understanding of the distributions) whereas panel (b) shows the same data (though
    extending to higher $v$) on a double-log scale. Panels (c) and (d) are $1-C(v)$ which
    is the fraction of particles with velocity $>v$.  Panels (e) and (f) show $S(v)$ from
    the cumulative dissipation with the vertical dashed line marking $v_{50}$ which is at
    50\% of the dissipation. It is clear that a fair part of the dissipation is from
    particles with velocities far out in the tail of the distribution.}
  \label{fig:vhist-8434-e100}
\end{figure*}

\section{Velocity distribution on linear and logarithmic scales}
\label{sec:linlog}

As jamming is approached the velocity distribution develops a wide tail and it then
becomes convenient to plot data on a double-log scale. The obvious drawback is that the
figures then become difficult to interpret and we therefore show a typical example of
$\cP(v)$ ---here obtained at $\phi=0.8434\approx\phi_J$ and $\gdot=10^{-7}$---in
\Fig{vhist-8434-e100}(a) and (b) plotted in two different ways with linear and logarithmic
scales.  \Fig{vhist-8434-e100}(a) shows that $\cP(v)$ has a peak at the low velocity
$v_p\approx8.6\times10^{-7}$ and from \Fig{vhist-8434-e100}(b), which is the same data
(though extending to higher $v$) on a double-log scale, it is clear that the distribution
extends up to much larger velocities, even above $100\;v_p$. \Fig{vhist-8434-e100}(c) and
(d) show $1-C(v)$, which is the fraction of particles with velocity $>v$. Here
$C(v)=\int_0^v\cP(v') dv'$ is the cumulative velocity distribution.

\Fig{vhist-8434-e100}(e) and (f) show the relative contribution to the shear viscosity for
particles with nonaffine velocity $<v$, obtained as $S(v)/\sigma$, and it is clear that a
fair part of the dissipation is from velocities far out in the tail of the
distribution. From the figure it follows that more than 25\% of the dissipation is for
$v>10^{-5}$ even though it could seem from \Fig{vhist-8434-e100}(a) that $\cP(v)$ is
negligible in that region and the same figure gives at hand that 50\% of the energy is
dissipated by only about 3.6\% of the fastest particles. This is, furthermore, a fraction
that keeps decreasing as $\gdot\to0$.

\section{On the origin of the wide velocity distribution}
\label{sec:wide}

A possible view on the anomalously large velocities that make up the tail of the velocity
distribution is that they occur when, due to a fluctuation, the critical volume fraction
for a particular configuration is anomalously small, so that the large velocities actually
reflect the elasto-plastic type behavior of a jammed configuration, rather than the
behavior of a packing of hard particles at constant pressure, below jamming.

That kind of picture is a natural one when approaching the subject from the analysis of
static packings. Quite a few things are however different in shear-driven simulations
close to $\phi_J$ and one of these is that it is not obvious that $p$ may be used to tell
about the ``true distance to jamming'', when the shearing systems are very far from
equilibrium.

In shear-driven jamming at low shear strain rates and well below the jamming density
$\phi_J\approx 0.8434$, say $\gdot=10^{-7}$ and $\phi=0.83$, things are simple. When
stopping the shearing and relaxing a configuration to a zero-energy state, the contact
number $z$ of the zero-energy state, is strongly correlated to $p$ of the initial
configuration. If one then tried to determine $\phi_c$ by compressing the relaxed
configuration further, one would presumably also find this $\phi_c$ to be strongly
correlated to $p$ of the initial configuration.

Closer to $\phi_J$---which is the region for most of our simulations---the correlation
between $p$ and $z$, however, becomes much smaller and the obvious reason is that the
relaxations often require substantial reorganizations and during these reorganizations the
system loses memory of it original state. A consequence is that we can no longer expect
$p$ to determine $\phi_c$.

It should also be noted that the fluctuations of $p$ are quite small. For $N=65536$
particles at $\phi=0.8434\approx\phi_J$, and shear strain rate $\dot\gamma=10^{-7}$ the
standard deviation of $p$ is, in relative terms, $\mathrm{std}(p)/p \approx 0.05$ and this
is by itself evidence that the fluctuations in $p$ cannot be the reason for the wide
velocity distribution.


%

\end{document}